%% file: ad-privacy.tex
\documentclass[sigplan,10pt,natbib=false]{acmart}
\acmYear{2024}\copyrightyear{2024}
\setcopyright{rightsretained}
\acmConference[SOSP '24]{ACM SIGOPS 30th Symposium on Operating Systems Principles}{November 4--6, 2024}{Austin, TX, USA}
\acmBooktitle{ACM SIGOPS 30th Symposium on Operating Systems Principles (SOSP '24), November 4--6, 2024, Austin, TX, USA}
\acmDOI{10.1145/3694715.3695965}
\acmISBN{979-8-4007-1251-7/24/11}

\input{style/boilerplate}
\input{style/macros}

\title{\sysname: Efficient On-device Budgeting for Differentially-Private Ad-Measurement Systems}
\author{Pierre Tholoniat*}
\affiliation{\institution{Columbia University} \country{}}
\author{Kelly Kostopoulou*}
\affiliation{\institution{Columbia University} \country{}}
\author{Peter McNeely}
\affiliation{\institution{Columbia University} \country{}}
\author{Prabhpreet Singh Sodhi}
\affiliation{\institution{Columbia University} \country{}}
\author{Anirudh Varanasi}
\affiliation{\institution{Columbia University} \country{}}
\author{Benjamin Case}
\affiliation{\institution{Meta Platforms, Inc.} \country{}}
\author{Asaf Cidon}
\affiliation{\institution{Columbia University} \country{}}
\author{Roxana Geambasu}
\affiliation{\institution{Columbia University} \country{}}
\author{Mathias L\'ecuyer}
\affiliation{\institution{University of British Columbia} \country{}}

\begin{document}
\date{}

\input{sections/00-abstract}

    

\settopmatter{printfolios=true}
\maketitle
\def\thefootnote{*}\footnotetext{These authors contributed equally to this work.
}\def\thefootnote{\arabic{footnote}}
\pagestyle{plain}

\input{sections/01-introduction}                             %
\input{sections/02-systematization-of-ad-measurement-apis}   %
\input{sections/03-system-overview}                          %
\input{sections/04-formal-modeling-and-analysis}             %
\input{sections/05-prototype}                                
\input{sections/06-evaluation}                               %
\input{sections/07-related-work}                             %
\input{sections/08-conclusion}
\printbibliography

\clearpage
\appendix
\noindent {\em Note: This appendix has not been peer-reviewed.}
\input{appendix/01-additional-use-cases}

\input{appendix/o6-performance-evaluation}
\input{appendix/02-system-algorithm}

\input{appendix/03-privacy-guarantees-proofs}
\input{appendix/04-idp-optimizations-proofs}

\input{appendix/05-idp-bias-detection}

\end{document}

%% file: style/boilerplate.tex
\usepackage{times}
\usepackage{graphicx}
\usepackage[export]{adjustbox}
\usepackage{array} %
\usepackage{color}
\usepackage{listings} %
\usepackage{xspace} %
\usepackage[noend]{algpseudocode} %
\usepackage{algorithm} %
\usepackage{wrapfig} %
\usepackage{flushend} %
\usepackage{multirow} %
\usepackage{booktabs} %
\usepackage{anyfontsize} %
\usepackage{paralist} %
\usepackage[font=footnotesize]{caption}
\usepackage{courier} %
\usepackage{verbatim}
\usepackage{calc}
\usepackage{amsthm}
\usepackage{pbox}
\usepackage{dsfont} %
\usepackage{soul} %
\usepackage{tabularx} %
\usepackage{titlesec}

\usepackage[justification=centering]{subfig}

\usepackage[
    style=numeric,
    sorting=nty,
]{biblatex}

\usepackage{tikz}
\usepackage{listings} %

\usepackage{pythonhighlight}

\definecolor{darkgreen}{rgb}{0,0.5,0}
\titleformat{\section}[block]
  {\normalfont\Large\bfseries} %
  {\thesection}{1em} %
  {} %
  [] %

\titleformat{\subsection}[block]
  {\normalfont\large\bfseries} %
  {\thesubsection}{1em} %
  {} %
  [] %
  
\titleformat{\subsubsection}[runin]
  {\normalfont\bfseries}
  {\thesubsubsection}{1em}
  {}[\newline]

%% file: style/macros.tex
\def\sysname{Cookie Monster\xspace}

\def\IPA{IPA\xspace}
\def\ARA{ARA\xspace}
\def\PAM{PAM\xspace}

\def\Hybrid{Hybrid\xspace}

\def\listOfAPIsWithCompanyNames{Meta and Mozilla's \IPA, Google's \ARA, Apple's \PAM, and Meta and Mozilla's \Hybrid}
\def\listOfAPIs{\IPA, \ARA, \PAM, and \Hybrid}

\def\baselineIPA{IPA-like\xspace}
\def\baselineARA{ARA-like\xspace}

\def\individualdp{individual DP\xspace}

\def\IDP{IDP\xspace}

\def\microbenchmark{microbenchmark\xspace}
\def\patcg{PATCG\xspace}
\def\criteo{Criteo\xspace}
\def\criteoplus{Criteo++\xspace}

\newcommand{\changeone}{Change One\xspace}

\newcommand{\ie}{{\em i.e.,~}}
\newcommand{\eg}{{\em e.g.,~}}
\newcommand{\aka}{{\em a.k.a.~}}

\newcommand\mydots{\makebox[.8em][c]{.\hfil.\hfil.}}

\def\Def{Def.~}
\def\Fig{Fig.~}

\def\Alg{Alg.~}
\def\Thm{Thm.~}
\def\Eq{Eq.~}

\newcommand{\heading}[1]{{\vspace{1pt}\noindent\bf{#1}}} %

\newtheorem{definition}{Definition}
\newtheorem{theorem}{Theorem}

\newcommand{\N}{\mathbb{N}}
\newcommand{\Z}{\mathbb{Z}}
\newcommand{\R}{\mathbb{R}}
\newcommand{\D}{\mathbb{D}}
\newcommand{\E}{\mathbb{E}}

\newcommand{\cA}{\mathcal{A}}
\newcommand{\cB}{\mathcal{B}}
\newcommand{\cC}{\mathcal{C}}
\newcommand{\cD}{\mathcal{D}}
\newcommand{\cE}{\mathcal{E}}
\newcommand{\cF}{\mathcal{F}}

\newcommand{\cI}{\mathcal{I}}

\newcommand{\cL}{\mathcal{L}}

\newcommand{\cM}{\mathcal{M}}
\newcommand{\cP}{\mathcal{P}}

\newcommand{\cU}{\mathcal{U}}

\newcommand{\cX}{\mathcal{X}}

\newcommand{\lap}{\operatorname{Lap}}

\algblock{Input}{EndInput}
\algnotext{EndInput}
\algblock{Output}{EndOutput}
\algnotext{EndOutput}
\algblock{Config}{EndConfig}
\algnotext{EndConfig}

\newcommand{\graycomment}[1]{\textcolor{gray}{// \em #1}}
\newcommand{\computebudget}{\operatorname{ComputeIndividualBudget}}

\def\enableComments{1}  
\ifnum\enableComments=1
\newcommand{\asaf}[1]{\textcolor{red}{\{Asaf: #1\}}}
\newcommand{\ml}[1]{\textcolor{purple}{\{Mathias: #1\}}}
\newcommand{\pierre}[1]{{\color{blue}{Pierre: #1}}}
\newcommand{\ben}[1]{{\color{darkgreen}{Ben: #1}}}
\newcommand{\kelly}[1]{{\color{violet}{Kelly: #1}}}
\newcommand{\roxana}[1]{{\color{cyan}{Roxana: #1}}}
\newcommand{\pierrerm}[1]{{\color{blue}{\st{#1}}}}
\newcommand{\pierrerp}[2]{{\color{blue}{\st{#1}{#2}}}}
\else
\newcommand{\asaf}[1]{}
\newcommand{\ml}[1]{}
\newcommand{\pierre}[1]{}
\newcommand{\ben}[1]{}
\newcommand{\kelly}[1]{}
\newcommand{\roxana}[1]{}
\newcommand{\pierrerm}[1]{}
\newcommand{\pierrerp}[2]{}
\fi

%% file: sections/00-abstract.tex
\begin{abstract}
With the impending removal of third-party cookies from major browsers and the introduction of new privacy-preserving advertising APIs, the research community has a timely opportunity to assist industry in qualitatively improving the Web's privacy. This paper discusses our efforts, within a W3C community group, to enhance existing privacy-preserving advertising measurement APIs.  We analyze designs from Google, Apple, Meta and Mozilla, and augment them with a more rigorous and efficient differential privacy (DP) budgeting component. Our approach, called {\em \sysname}, enforces well-defined DP guarantees and enables advertisers to conduct more private measurement queries accurately. By framing the privacy guarantee in terms of an individual form of DP, we can make DP budgeting more efficient than in current systems that use a traditional DP definition.  We incorporate \sysname into Chrome and evaluate it on microbenchmarks and advertising datasets. Across workloads, \sysname significantly outperforms baselines in enabling more advertising measurements under comparable DP protection.
\end{abstract}

%% file: sections/01-introduction.tex
\section{Introduction}
\label{sec:introduction}

Web advertising is undergoing significant changes, presenting a major opportunity to enhance online privacy. For years, numerous entities, often without users' knowledge, have exploited Web protocol vulnerabilities, such as third-party cookies and remote fingerprinting, to track user activity across the Web. This data has been used to target individuals with ads and assess ad campaign performance. Two key shifts are reshaping this landscape. First, major browsers are making it more difficult to track users across websites. Apple’s Safari and Mozilla’s Firefox blocked third-party cookies in 2019~\cite{safari-disables-cookies} and 2021~\cite{firefox-disables-cookies}, respectively, while Google Chrome will soon facilitate users' choice of disabling these cookies~\cite{new-google-cookies-announcement}. Additionally, browsers are strengthening defenses against IP tracking~\cite{apple-private-relay} and remote fingerprinting~\cite{firefox-disables-cookies,disable-remote-fingerprinting-apple,disable-remote-fingerprinting-google}.

Second, acknowledging the critical role online advertising plays in the Web economy -- and the impossibility of perfect tracking protection -- browsers are introducing explicit APIs to measure ad effectiveness and enhance ad delivery while protecting individual privacy.  Early designs, like Apple’s PCM~\cite{PPACAblog2019} and Google’s FLoC~\cite{floc}, focused on intuitive but not rigorous privacy methods, resulting in limited adoption due to poor utility~\cite{pcm-poor-utility} or privacy~\cite{floc-poor-privacy}. Recently, browsers have shifted to theoretically-sound privacy technologies -- such as differential privacy (DP), secure multi-party computation (MPC), and trusted execution environments (TEEs) -- in the hope of achieving better privacy-utility tradeoffs.

However, substantial challenges remain in implementing these privacy technologies at Web scale. The research community now has a timely opportunity -- and responsibility -- to assist industry in refining these technologies to deliver both strong privacy protections and meet advertising needs. Only by addressing these challenges can we hope to drive adoption of privacy-preserving APIs, remove incentives for individual tracking, and meaningfully improve Web privacy.

This paper focuses on our efforts to analyze and enhance current {\em ad-measurement APIs} (a.k.a., attribution-measurement APIs), which enable advertisers to measure and optimize the effectiveness of their ad campaigns based on how often people who view or click certain ads go on to purchase the advertised product. While separate {\em ad-targeting APIs} are also under development~\cite{protected-audience-api}, we concentrate on {\em ad-measurement APIs}.

The W3C's Private Advertising Technology Community Group (PATCG)~\cite{patcg} is working towards an interoperable standard for private ad-measurement APIs. Leading proposals include Google’s Attribution Reporting API (ARA)~\cite{ara}, Meta and Mozilla’s Interoperable Private Attribution (IPA)~\cite{ipa}, Apple’s Private Ad Measurement (PAM)~\cite{pam}, and a hybrid proposal~\cite{hybrid-proposal}. Our first contribution is a systematization of these proposals into abstract models, followed by a comparative analysis to identify opportunities for improving their privacy-utility tradeoffs (\S\ref{sec:existing-ad-measurement-systems}).

We focus on the differential privacy (DP) component, present in all four systems. DP is used to ensure advertisers cannot learn too much about any single user through measurement queries. Each system employs a {\em privacy loss budget}, accounting for the privacy loss incurred by each query. Once the budget is exhausted, further queries are blocked. This process, called {\em DP budgeting}, is handled centrally in IPA, but in the other systems, DP budgeting is done separately by each device. We observe that this {\em on-device budgeting} cannot be formalized under standard DP and instead requires a variant, {\em \individualdp} (\IDP) or personalized DP~\cite{popl15}, for proper formalization. Our formal modeling and analysis of on-device budgeting under \IDP form our second contribution (\S\ref{sec:formal-model-and-analysis}).

Through our \IDP formalization, we uncover optimizations that enhance utility in on-device budgeting systems, allowing advertisers to execute more accurate queries under the same DP budget. \IDP enables devices to maintain their own, separate DP guarantees and to account for privacy loss based on the device’s data. This lets a device deduct zero privacy loss if it lacks relevant data for a query. Notably, one such optimization is already used in ARA, though without formal justification. Our third contribution is providing formal proof for this optimization as well as other, novel optimizations that can further improve the privacy-utility tradeoff.

Our final contribution is a prototype implementation of our optimized DP budgeting system, called {\em \sysname}, integrated into ARA within Chrome (\S\ref{sec:system-overview}, \S\ref{sec:prototype}). 
\sysname is the first ad-measurement system to enforce a fixed, user-time DP guarantee~\cite{KMRTZ20Guidelines}, improving on the event-level guarantees of ARA.
We evaluate \sysname on microbenchmarks and advertising datasets (\S\ref{sec:evaluation}), showing that it delivers $\times$1.16--2.88 better query accuracy compared to a user-time version of ARA and substantially outperforms IPA, which exhausts its budget very early.  Our prototype is available at \url{https://github.com/columbia/cookiemonster} and has been incorporated into a W3C draft report on privacy-preserving attribution from Mozilla~\cite{https://private-attribution.github.io/api/}.

%% file: sections/02-systematization-of-ad-measurement-apis.tex
\vspace{-0.5em}\section{Review of Ad-Measurement APIs}
\label{sec:existing-ad-measurement-systems}

We review the designs of privacy-preserving ad-measurement systems considered for a potential interoperable standard at PATCG: \listOfAPIsWithCompanyNames. \ARA and \IPA are implemented; \PAM and \Hybrid exist only as design docs. We abstract their functionality for comparison and articulate the improvement opportunity addressed in this paper.

\vspace{-0.5em}\subsection{Example Scenario}
\label{sec:running-example}

We use a fictitious scenario to illustrate the motivation and requirements of ad-measurement systems from two key perspectives: Ann, a web user, and Nike, an advertiser measuring ad campaign effectiveness. While real-world players like first-party ad platforms (e.g., Meta) and ad-techs (e.g., Criteo) typically run measurement queries on behalf of advertisers, for simplicity, we assume the advertiser performs its own measurements. We discuss the other players in Appendix~\ref{sec:appendix:additional-use-cases}.

\heading{User perspective.} Ann visits various {\em publisher} sites, such as nytimes.com and facebook.com, where she sees ads. She understands that ads fund the free content she enjoys and occasionally finds them useful, like when she clicked on a Nike ad for running shoes on nytimes.com and later purchased a pair. However, Ann values her privacy and expects {\em no cross-site tracking}, meaning no site should track her across different websites. She also expects {\em limited within-site linkability}, preventing even a single site from linking her activities across cookie-clearing browsing sessions (e.g., incognito sessions). Ann accepts that some privacy loss is necessary for effective advertising but expects it to be {\em explicitly bounded} and {\em transparently reported} by her browser.

\begin{figure}[t]
	\centering
	\includegraphics[width=.7\linewidth, frame]{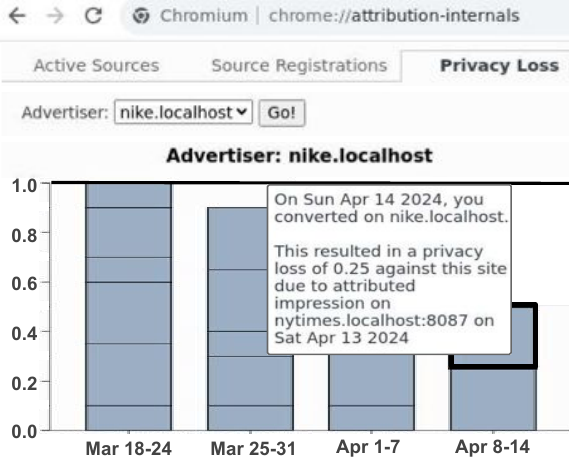}
	\caption{{\bf Privacy loss dashboard.} Screenshot from our Chrome implementation of \sysname (minimally edited for visibility).}
 \vspace{-0.5em}
	\label{fig:privacy-loss-dashboard-screenshot}
\end{figure}

\Fig\ref{fig:privacy-loss-dashboard-screenshot} shows a screenshot of the privacy loss dashboard we developed for \sysname in Chrome, where Ann can monitor the privacy loss resulting from various sites and intermediaries querying her ad interactions, including {\em impressions} (\eg ad views and clicks) and {\em conversions} (\eg purchases, cart additions). While Ann may not grasp the concept of differential privacy that underpins the reported privacy loss, she trusts her browser to always enforce protective bounds on it.

\heading{Advertiser perspective.}
Nike runs multiple ad campaigns for its running shoes, some emphasizing shock-absorbing technology, others focusing on aesthetics. Nike seeks to understand which campaigns perform best across different demographics and contexts (e.g., publisher sites, content types). In the past, Nike used third-party cookies and device fingerprinting\footnote{The example is fictitious, as are claims regarding the companies mentioned.} to track individuals from ad impressions to purchases, attributing purchase value using an {\em attribution function}, such as last-touch (giving all credit to the last impression) or equal credit (splitting value among recent impressions). Using such {\em attribution reports} from many users, Nike measured the purchase value attributed to different campaigns and optimized future ad targeting.

Now that third-party cookies are disabled on multiple browsers and fingerprinting is harder, Nike is transitioning to ad-measurement APIs, expecting similar attribution measurements with comparable accuracy. Nike understands that ad measurement has always involved some imprecision (e.g., due to cookie clearing or fraud), so its expectation of accuracy from these APIs is not stringent. Nike plans to conduct numerous attribution measurements over time to adjust to changing user preferences and product offerings. These measurements are single-advertiser summation queries, a key query type that ad-measurement systems aim to support.

\begin{figure*}[t]
    \centering
    \includegraphics[width=\linewidth]{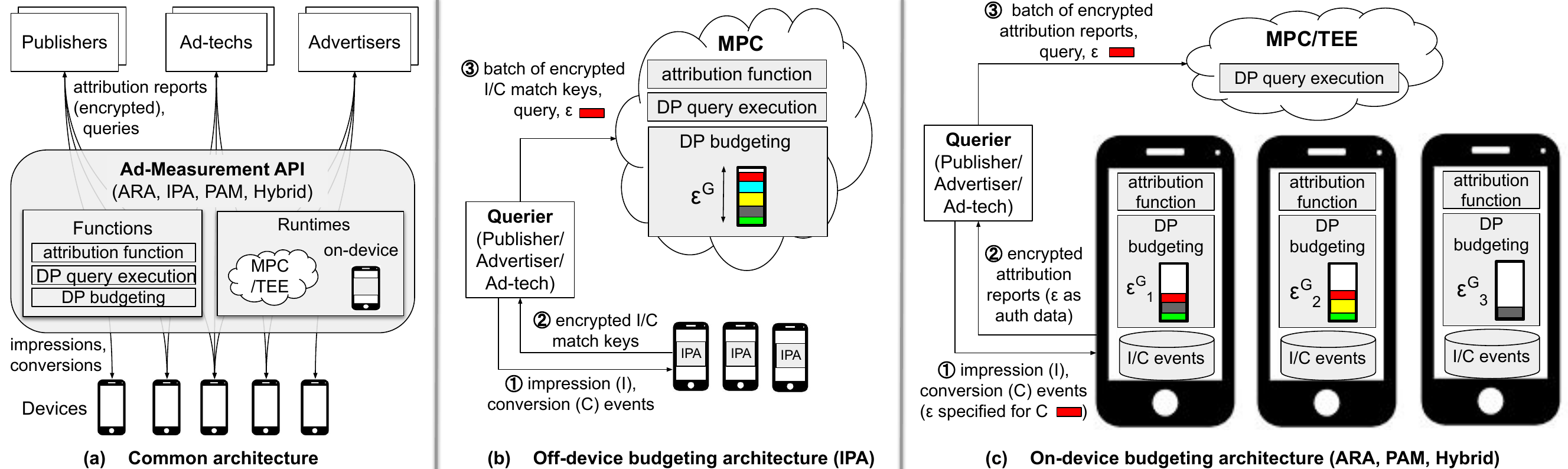}
    \caption{{\bf Architectures of ad-measurement systems.} Common structure, with a key difference in where attribution and DP budgeting occur: off-device (\IPA) vs. on-device (\ARA, \PAM, \Hybrid).}
    \label{fig:existing-ad-measurement-systems}
\end{figure*}

\vspace{-0.5em}\subsection{Ad-Measurement Systems}
\label{sec:apis-overview}

\listOfAPIs aim to balance user privacy with utility for advertisers and other Web-advertising parties (referred to as {\em queriers}). Utility is defined as the number of accurate measurement queries a querier can execute under a privacy constraint. Despite variations in terminology, privacy properties, and mechanisms, these systems share key similarities. A commonality is the use of DP techniques, with \ARA focusing on event-level DP, while \IPA, \PAM, and \Hybrid emphasize user-time DP. This paper focuses on user-time DP, applied per querier site, as defined in \S\ref{sec:individual-dp}.

\heading{Common architecture.}
The high-level architecture of all four systems is similar (see \Fig\ref{fig:existing-ad-measurement-systems}a). All systems act as intermediaries between user devices and sites. Previously, these parties collected impression and conversion events directly, matched them through third-party cookies, performed attribution, and aggregated reports. To break these privacy-infringing direct data flows, ad-measurement systems interpose a DP querying interface over impression and conversion data.

All systems include three core components: (1) the {\em attribution function}, which matches conversions to relevant impressions on the same device and assigns conversion value to impressions based on an attribution logic like last-touch; (2) {\em DP query execution}, which aggregates reports and adds noise for DP guarantees; and (3) {\em DP budgeting}, which tracks privacy loss from each query using DP composition and enforces a maximum on total privacy loss, called a {\em DP budget}.

A key difference is where these components are executed. In \IPA, all components run off-device within an MPC involving multiple helper servers. In \ARA, \PAM, and \Hybrid, attribution and DP budgeting occur on-device, while DP query execution is off-device, in an MPC (\PAM, \Hybrid) or TEE (\ARA). The MPC/TEE is trusted not to leak inputs, and the devices are trusted to safeguard their own data. The placement of attribution and DP budgeting is crucial for this paper.

\heading{Off-device budgeting (\IPA).}
\Fig\ref{fig:existing-ad-measurement-systems}b illustrates \IPA, which operates in a standard centralized-DP setting. The MPC handles all three functions, while the device's role is limited to generating a {\em match key} to link impressions and conversions. For example, when nytimes.com sends an ad for Nike shoes to Ann’s device~\textcircled{1}, the device responds with a match key, secret-shared and encrypted toward the MPC helper servers~\textcircled{2}. When Ann later purchases the shoes on nike.com, her device sends the same key to the MPC, also secret shared and encrypted toward the helpers.  Periodically, NYtimes sends batches of encrypted impression match keys to Nike, who cannot directly match these with its conversion match keys due to the encryption and secret sharing. Instead, Nike collects its conversion match keys and NYtimes' impression match keys into batches and submits them to the MPC, specifying the privacy budget $\epsilon$ to spend on the query~\textcircled{3}. The MPC checks the budget, matches impressions to conversions, applies the attribution function with an $L^1$ cap for sensitivity control, aggregates the data, and adds DP noise to enforce $\epsilon$-DP. The MPC tracks and deducts Nike's privacy budget, refusing further queries once the budget is exhausted until the per-site budget is ``refreshed'' (e.g., daily).

\heading{On-device budgeting (\ARA, \PAM, \Hybrid).}
\Fig\ref{fig:existing-ad-measurement-systems}c shows the on-device architecture, which operates in a rather non-standard DP setting.  While DP query execution occurs centrally on the MPC or TEE, attribution and DP budgeting are done {\em separately on each device}. Every device maintains a timeseries database of impression and conversion events. When Ann sees an ad for Nike on nytimes.com, her device records it locally~\textcircled{1}. Later, when she buys shoes on nike.com, Nike requests an attribution report from her device. Ann’s device checks its database for relevant impressions, applies the attribution function with an $L^1$ cap, and sends an {\em attribution report}~\textcircled{2}, either secret-shared and encrypted toward the helper parties (for MPC) or directly encrypted to a TEE. Nike aggregates attribution reports from multiple users, submits them to the MPC or TEE, which performs DP aggregation, adding noise based on Nike’s $\epsilon$ parameter~\textcircled{3}. The MPC/TEE ensures each report is used only once for sensitivity control.

DP budgeting in on-device systems differs from centralized DP by accounting for privacy loss when the advertiser requests a conversion report, prior to query execution. When Nike requests a report, it specifies the $\epsilon$ parameter for the future query. The device checks Nike's budget locally, generates and encrypts the report (with secret sharing if MPC is used), includes $\epsilon$ as authenticated data, and deducts $\epsilon$ from Nike's local budget. Since the budget is spent at the device, each report can only be used once, so the device includes a unique nonce with every report in authenticated data and the MPC/TEE tracks report nonces to prevent reuse.

\heading{Threat models.}
The threat models differ based on whether an MPC or TEE is used. In all cases, MPC/TEE systems are trusted to protect inputs and intermediate states. For MPC, the deployment models assume either a three-party, malicious, honest-majority MPC protocol (\IPA, Hybrid)~\cite{ipa} or a two-party malicious protocol (\PAM). The querier selects MPC parties from a browser-configured list, typically relatively trusted Web organizations like Cloudflare. The device secret shares the report and encrypts it toward the chosen parties after report generation.

\vspace{-0.5em}\subsection{Improvement Opportunity}
\label{sec:opportunity}

On-device budgeting systems offer certain advantages over off-device systems but also present a key challenge, which we aim to address. First, on-device systems can enhance user transparency by putting the user's device in control of per-site budgets and the tracking of privacy losses incurred by the user due to specific attribution reports the device releases to various querier sites, as seen in the \sysname privacy loss dashboard (\Fig\ref{fig:privacy-loss-dashboard-screenshot}). In contrast, in \IPA, the device can only track the encrypted match keys returned by the device, not the specific privacy losses users incur through subsequent matching and aggregation in the MPC.

Second, on-device systems allow for finer-grained budgeting. While off-device systems enforce a global site-wide budget $\epsilon^G$, on-device systems maintain a per-device budget $\epsilon_d^G$, which is only consumed for queries involving that device. This granularity enables Nike, for instance, to continue querying other users' reports even if it exhausts Ann’s budget. However, this behavior requires formalization under the less standard (but equally protective) privacy definition known as individual DP (IDP)~\cite{popl15}, which allows enforcement of a separate privacy guarantee for each device.

The challenge lies in formalizing the data, query, and system model that capture the behavior of on-device ad-measurement systems, and in proving its IDP properties.  This formalization then opens opportunities for further optimizing DP budgeting in on-device systems by deducting privacy loss based on the device's data. However, it also requires keeping the remaining privacy budgets on each device private, as revealing these budgets leaks data. This paper presents a formally-justified, practical and efficient DP budgeting module, {\em \sysname}, designed for on-device systems like \ARA, \PAM, and \Hybrid, which maximizes utility while maintaining DP guarantees.

%% file: sections/03-system-overview.tex
\vspace{-0.5em}\section{\sysname Overview}
\label{sec:system-overview}

The design of \sysname is guided by three principles. First, it must enforce well-defined DP guarantees at an industry-accepted granularity. We adopt a fixed ``user-time'' DP guarantee for each querier, supported by \IPA, \PAM, and \Hybrid, and recognized by Apple, Meta, and Mozilla as the minimum acceptable. Second, \sysname must support similar use cases and queries as existing systems. We focus on the single-advertiser measurement query from \S\ref{sec:running-example}, though we briefly discuss in Appendix~\ref{sec:appendix:additional-use-cases} how a multi-advertiser optimization query might apply. Finally, \sysname must not introduce new vectors for illicit tracking, given increasing browser efforts to prevent tracking both across sites and within-site across cookie refreshes.

\begin{figure}[t]
    \centering
    \includegraphics[width=\linewidth]{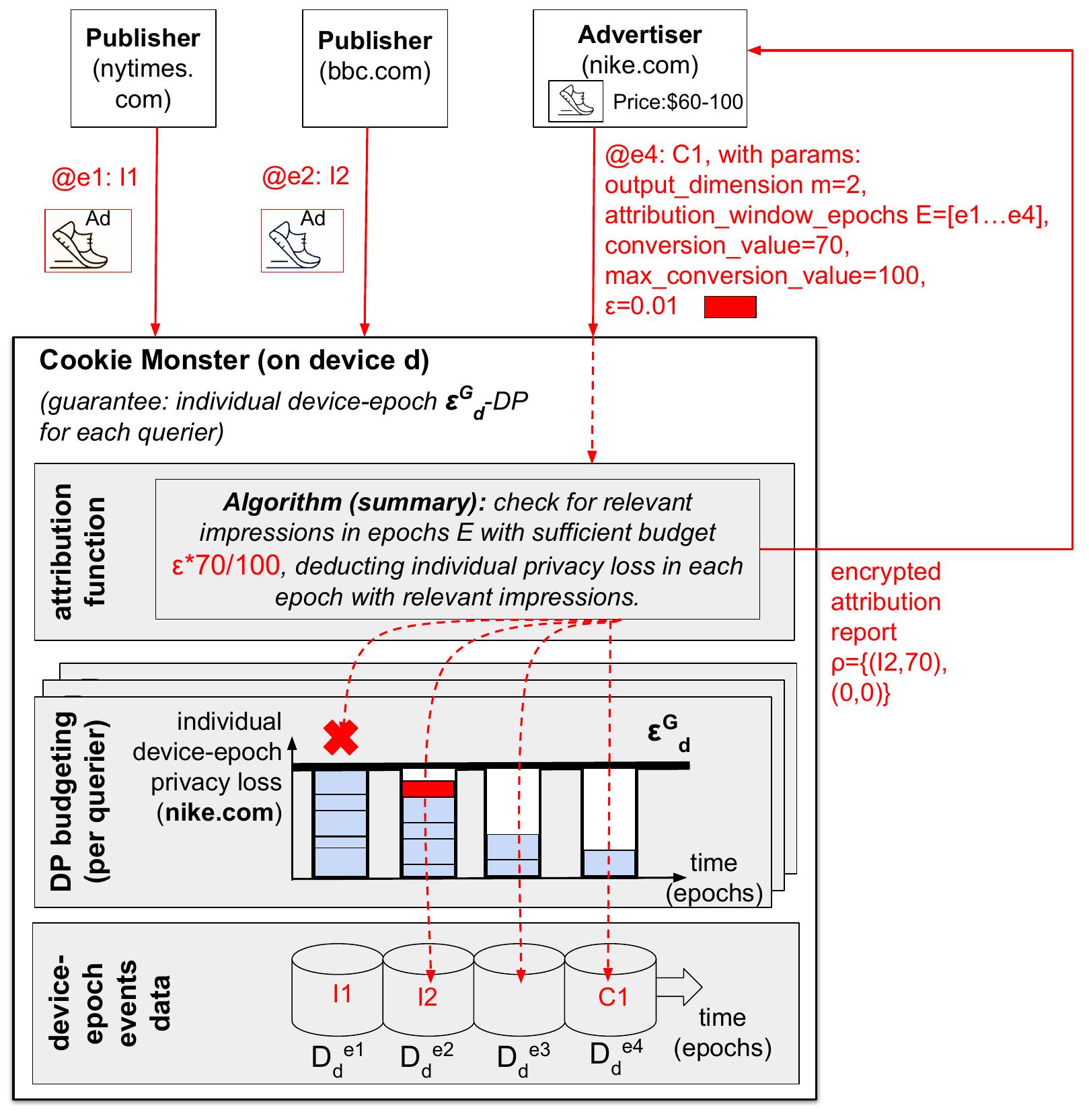}
    \caption{{\bf \sysname architecture and example execution (red overlay).}
    \S\ref{sec:architecture} describes the architecture and \S\ref{sec:execution-example} the example execution. Notation: $@e_1:I_1$ indicates that Ann's device receives an impression $I_1$ of a Nike shoe ad from nytimes.com in epoch $e_1$. Red dotted arrows show the attribution function's search for impressions over epochs $e_1-e_4$.
    }
    \label{fig:architecture}
\end{figure}

\Fig~\ref{fig:architecture} presents \sysname's architecture with an example execution overlaid. We describe each aspect below.

\vspace{-0.5em}\subsection{Architecture}
\label{sec:architecture}

\sysname adopts on-device budgeting, similar to \ARA, \PAM, and \Hybrid. DP query execution occurs off device, in an MPC or TEE, trusted not to leak inputs or intermediate states. Since \sysname does not modify this component, it is omitted from \Fig\ref{fig:architecture}; we think of it as a trusted {\em aggregation service}. \sysname modifies the on-device component, based on \ARA in our prototype. While the external APIs remain unchanged, we modify: (1) the on-device events database to support a ``user-time'' guarantee, and (2) the internals of the attribution function and DP budgeting to enforce this guarantee efficiently.

\sysname enforces {\em individual device-epoch $\epsilon^G_d$-DP} for each querier site, formally defined in \S\ref{sec:individual-dp}. This device-epoch granularity aligns with traditional ``user-time'' from DP literature~\cite{KMRTZ20Guidelines, sage, privatekube}, though we rename it to reflect that a user's complete activity is not directly observable by a device or browser, the scope in which \sysname operates. We partition the on-device events database into time-based {\em epochs}, such as weeks or months. In each epoch $e$, device $d$ collects impression and conversion events into a {\em device-epoch database} $D_d^e$. Queriers submit multiple queries over time, accessing data from one or more epochs. For each epoch $e$, \sysname ensures that no querier learns more about device $d$'s data in $e$ than permitted by an $\epsilon^G_d$-DP guarantee.

The {\em DP budgeting} in \sysname is implemented using privacy filters~\cite{RRUV16}, which ensure that the cumulative privacy loss from a series of queries does not exceed a pre-specified budget. For each querier, \sysname maintains multiple filters -- one for each device-epoch database. \Fig~\ref{fig:architecture} shows these filters for nike.com. Each filter is initialized with a privacy budget $\epsilon^G_d$ and monitors cumulative privacy loss for queries involving data from that epoch.

In on-device systems, privacy loss is accounted for when the attribution report is generated, not when the query is executed. The {\em attribution function} is responsible for generating these reports. Upon a conversion, the function checks for relevant impressions in the device-epoch databases within a specified attribution window. Privacy filters prevent use of impression data from epochs with insufficient budget.

For epochs with sufficient budget, the filter allows access to the device-epoch data and deducts privacy loss. Under standard centralized DP, this loss would be $\epsilon$, the DP parameter enforced later by the MPC or TEE during aggregation. However, our theoretical analysis of on-device budgeting reveals that viewing the system under an individual-DP lens opens opportunities to optimize privacy accounting, often allowing deductions of ``less than $\epsilon$.'' \S\ref{sec:formal-model-and-analysis} outlines our theoretical analysis, a major contribution in this paper. We dedicate the remainder of this section to providing the systems view of our theory, including an execution example (\S\ref{sec:execution-example}), \sysname's algorithm, which is backed by our theory (\S\ref{sec:algorithm}), and a discussion on mitigating IDP-induced bias (\S\ref{sec:idp-implications}).

\vspace{-0.5em}\subsection{Execution Example}
\label{sec:execution-example}

The red overlay in \Fig\ref{fig:architecture} illustrates the attribution function's operation for the example from \S\ref{sec:running-example}. Ann receives two impressions of Nike shoe ads: one in epoch $e_1$ and another in $e_2$, with no impressions in $e_3$. Later, in epoch $e_4$, Ann buys the shoes, and nike.com registers a conversion $C1$. It requests an attribution report with parameters: the set of epochs $E$ to search for impressions, the maximum number of impressions $m$ to attribute value to, the conversion value ($\$70$), and $\epsilon$, the privacy parameter enforced by the MPC or TEE when executing the aggregation query.

The shoes' price ranges by color, with a maximum of \$100. While Ann's conversion is \$70, Nike’s query will include conversions up to \$100. Thus, for a summation query with the Laplace mechanism, the noise added to the aggregate depends on $100/\epsilon$, where $100$ is the {\em global sensitivity} of the summation (i.e., the largest change {\em any} device-epoch can contribute). Ann, with a purchase of \$70, can only contribute up to \$70 across her device-epochs.

Here, IDP lets us optimize privacy loss based on {\em individual sensitivity}, the maximum change that {\em a specific} device-epoch can make on the query output. In this case, Ann's device only deducts $\epsilon' = \$70/\$100 * \epsilon$ from the privacy filters of the epochs in the attribution window $E$. This is one optimization enabled by IDP. Another is that if no relevant impressions exist in an epoch (e.g., $e_3$ in \Fig\ref{fig:architecture}), we need not deduct anything, since the individual sensitivity for that epoch is 0 and thus its privacy loss is also 0. \S\ref{sec:idp-optimizations} formalizes global and individual sensitivities and details further optimizations.

In \Fig\ref{fig:architecture}, \sysname's attribution function checks epochs $e_1-e_4$ for relevant impressions. In $e_1$, access to data $D_d^{e_1}$ is denied because the filter has exhausted nike.com's budget. In $e_2$, the filter allows access, and a relevant impression $I_2$ is found, deducting $\epsilon'$ (shown as a red square in the $e_2$ filter). In $e_3$, there is budget, but no relevant impression is found, so no deduction occurs. Finally, in $e_4$, where the conversion happened but no impression occurred, then through a formalization of publicly available information that we support (\S\ref{sec:formal-system-model}), we can justify that no privacy loss occurs in $e_4$.

The final attribution report assigns the \$70 value to the single impression $I_2$ and includes a null value for the second attribution, as Nike requested two. If no impressions were found, or Nike also ran out of budget in $e_2$, the attribution function would return a report with two null values to avoid leaking information about ad presence.

\vspace{-0.5em}\subsection{Algorithm}
\label{sec:algorithm}

Listing~\ref{algo_listing} shows how \sysname computes an attribution report. The {\small \tt compute\_attribution\_report} function receives an {\small \tt attribution\_request}, which encapsulates all querier-provided parameters, sanitized by the device.  Key parameters include: 
\begin{enumerate}
    \item the window of epochs to search for relevant events ({\small \tt epochs} parameter);
    \item the requested privacy budget ({\small \tt requested\_epsilon});
    \item logic for selecting relevant events ({\small \tt select\_relevant\newline\_events});
    \item the attribution policy, such as last-touch or equal-credit ({\small \tt compute\_attribution});
    \item two global sensitivity parameters: {\small \tt report\_global\newline\_sensitivity}, the maximum change a device-epoch can make to the output of the report generation function, and {\small \tt query\_global\_sensitivity}, the maximum across all devices and reports;
    \item p-norm, based on the DP mechanism in MPC/TEE, e.g., 1-norm for Laplace and 2-norm for Gaussian.
\end{enumerate}

All parameters follow a predefined protocol, and while the algorithm is general enough to handle different mechanisms and p-norm sensitivities, our DP result (Thm.~\ref{thm:e2e_idp}) focuses on pure DP, assuming the Laplace mechanism and $L_1$ sensitivity.

Computing an attribution report consists of four steps.

{\bf Step~1:} \sysname invokes the querier-provided \newline{\small \tt select\_relevant\_events} to select relevant events from each separate epoch in the attribution window, such as impressions with a specific campaign ID.

{\bf Step 2:} For each epoch, \sysname computes the individual privacy loss resulting from the querier's query, following the IDP optimizations in Thm.~\ref{thm:individual_sensitivity_of_reports_queries}. Three cases:
\begin{enumerate}
    \item if the epoch has no relevant events, privacy loss is zero;
    \item if a single epoch is considered, privacy loss is proportional to the $L_p$-norm of the attribution function output;
    \item if multiple epochs are considered, privacy loss is proportional to the report's global sensitivity.
\end{enumerate}
The privacy loss is scaled by {\small \tt requested\_epsilon} and the query's global sensitivity. In \S\ref{sec:execution-example}, the report's global sensitivity is $70$, and the query's global sensitivity is $100$.

{\bf Step 3:} For each epoch, we attempt to deduct the computed privacy loss from the querier's budget for that epoch, ensuring atomic, thread-safe checks. If the filter has sufficient budget, the epoch's events are used for attribution; otherwise, they are dropped. The justification for dropping contributions is provided in Theorem~\ref{thm:e2e_idp}.

{\bf Step 4:} The attribution function is applied across events from all epochs, following the querier's policy.  The device ensures that the attribution computation: (1) respects the querier's specified {\small \tt report\_global\_sensitivity} by clipping the attribution histogram to ensure its $L_p$-norm is $\le$ {\small \tt report\_global\_sensitivity}, and (2) produces encrypted outputs indistinguishable from others. For (2), the device ensures a fixed dimension for the attribution report by padding or dropping elements. For instance, if only one relevant impression is found but two are requested, the output vector is padded with a null entry.

\begin{lstlisting}[language=Python,caption={{\bf \sysname Algorithm}},captionpos=b,label={algo_listing}]
# Global variables: events_database, privacy_filters.
def compute_attribution_report(attribution_request):
  relevant_events_per_epoch = {}
  for epoch in attribution_request.epochs:
    relevant_events = attribution_request.select_relevant_events(events_database[epoch])  # Step 1
    individual_privacy_loss = compute_individual_privacy_loss(relevant_events, attribution_request)  # Step 2  
    filter_status = privacy_filters[attribution_request.querier_site][epoch].check_and_consume(individual_privacy_loss)  # Step 3
    if filter_status == "out_of_budget":
      relevant_events = {}
    relevant_events_per_epoch[epoch] = relevant_events
  return attribution_request.compute_attribution(relevant_events_per_epoch)  # Step 4

def compute_individual_privacy_loss(epoch_events, attribution_request):
  if epoch_events == {}:  # Case 1 in Theorem 4
    return 0
  if len(attribution_request.epochs) == 1:  # Case 2 in Theorem 4
    individual_sensitivity = attribution_request.pnorm(attribution_request.compute_attribution(relevant_events))
  else:  # Case 3 in Theorem 4
    individual_sensitivity = attribution_request.report_global_sensitivity
  return attribution_request.requested_epsilon * individual_sensitivity / attribution_request.query_global_sensitivity
\end{lstlisting}

For the example in \S\ref{sec:execution-example}, this algorithm is invoked with an {\small \tt attribution\_request} where {\small \tt querier\_site} = ``nike.com,'' {\small \tt epochs} $= [e_1-e_4]$, {\small \tt report\_global\_sensitivity} $= 70$, {\small \tt query\_global\_sensitivity} $= 100$. Function {\small \tt select\newline\_relevant\_events} filters impressions by campaign ID, {\small \tt pnorm} returns the L1-norm of the attribution histogram, and {\small \tt compute\_attribution} divides the conversion value of $70$ across at most two impressions, padding with nulls as needed. This attribution function has sensitivity $70$.

\subsection{Bias Implications of IDP}
\label{sec:idp-implications}

The execution example and algorithm demonstrate \sysname's budget savings, confirmed in Section \ref{sec:evaluation}, where we show that these savings allow more accurate queries than ARA and IPA under the same privacy guarantees. However, IDP can introduce bias into query results. Since privacy loss and remaining budgets depend on data, they must remain hidden from advertisers. When a device exhausts its budget for an epoch, it continues participating in queries with ``null'' data, protecting privacy but potentially introducing bias. For example, Nike's report should have included two impressions, but running out of budget in epoch $e_1$ meant $I_1$ wasn't returned, altering the report undetectably.

This bias is a general challenge for all systems operating on IDP, including all existing ad-measurement systems with on-device budgeting -- although this challenge is not always acknowledged or handled. Indeed, ARA incorporates code to send null reports when budgets are exhausted and its documentation states that these nulls must be sent to preserve privacy~\cite{ara-doc-null-reports}. 
 Such nulls would add bias to query results. In absence of proper IDP formulation, a rudimentary justification we have seen for sending nulls in on-device systems is to prevent revealing budget exhaustion, which could facilitate remote fingerprinting, a concern actively addressed by browsers.  Our paper reveals a deeper issue: these systems inherently operate under IDP, and IDP systems must keep budgets hidden, which can lead to bias.  Acknowledging this bias opens pathways to mitigate it.

Any (DP or IDP) system must tolerate some error. In ad measurement, high error tolerance is common due to factors like tracking inaccuracies and fraud. The goal is to equip queriers with tools that rigorously bound errors from both DP noise and IDP bias, allowing for informed decision-making. Previous work on centralized-budgeting IDP has developed methods to bound bias using global sensitivity~\cite{FZ21} and periodic DP counting queries~\cite{YKK+23, FZ21}. These approaches require adaptation to on-device budgeting, given the lack of centralized privacy-loss tracking and non-i.i.d. report sampling. We leave it for future work to develop advanced bias-management tools and here only present a rudimentary approach, which we implement in \sysname and evaluate in \S\ref{sec:evaluation:bias-detection} as a proof-of-concept that bias can be effectively managed in on-device budgeting systems.

Our approach adds a {\em side query} to each attribution query, which bounds potential error from out-of-budget epochs. With each report, the querier requests a boolean flag indicating whether the report could be affected by an out-of-budget epoch. This flag is bundled with the attribution report, secret-shared, and encrypted toward the MPC/TEE. The querier receives a DP-aggregated count of how many reports could be erroneous out of its total batch. With the count, the querier computes a high-probability upper bound on the error from both DP noise and IDP bias. The querier can then filter the results of its queries based on this error bound, ignoring those with unacceptable error.  Formalization and proof of this mechanism's correctness are deferred to Appendix~\ref{sec:appendix:bias-detection}.

Consider last-touch attribution. If no epoch in the attribution window is out of budget or an impression is found in a later epoch, the device returns a 0-valued error assessment, indicating no bias. If no impression is found in epochs later than the out-of-budget epoch, the device returns a 1-valued error assessment, signaling potential bias. This information is encrypted and only accessible to the querier after DP aggregation by the MPC/TEE.

This mechanism lets queriers manage IDP-induced error rigorously, though it consumes additional privacy budget. In Steps 3 and 4 of Listing~\ref{algo_listing}, each epoch that is not out of budget must deduct privacy loss for the side query. Fortunately, since the side query is a count query with lower sensitivity than the main query, \sysname's optimizations still provide benefits. 
Our evaluation shows that even with bias detection, \sysname consumes less privacy and incurs lower errors compared to ARA and IPA (\S\ref{sec:evaluation:bias-detection}).%

%% file: sections/04-formal-modeling-and-analysis.tex
\section{Formal Modeling and Analysis}
\label{sec:formal-model-and-analysis}

This section outlines the theoretical analysis behind \sysname's design, divided into three parts: \S\ref{sec:formal-system-model} introduces a formal model that captures the behavior of on-device budgeting systems, including \sysname but also \ARA and \PAM. \S\ref{sec:idp-formulation-and-guarantees} analyzes this model under IDP, proving that \sysname bounds cross-site leakage and within-site linkability. Finally, \S\ref{sec:idp-optimizations} details and justifies the optimizations enabled by IDP, both ones inherently employed in \ARA and new ones that our theory uncovers.

\input{sections/04.1-formal-system-model}

\input{sections/04.2-idp-formulation}
\input{sections/04.3-idp-optimizations}

%% file: sections/04.1-formal-system-model.tex
\subsection{Formal System Model}
\label{sec:formal-system-model}

To rigorously analyze privacy properties and identify optimization opportunities in on-device budgeting systems for ad measurement, we must establish a formal model of their behavior. Current ad-measurement systems lack such models, preventing formal analysis or justification of optimizations. Although our model is tailored to \sysname, it can also serve as a foundation for analyzing other systems.

We define the data and queries \sysname operates on, from the perspective of a fixed querier (e.g., advertiser, publisher, or ad-tech). Appendix~\S\ref{sec:appendix:e2e_algorithm} formalizes the end-to-end algorithm, incorporating these models and the \sysname behavior outlined in \S\ref{sec:system-overview}. Since this algorithm is used solely to prove the DP guarantees in \S\ref{sec:idp-formulation-and-guarantees}, we omit it here.

\vspace{-0.5em}\subsubsection{Data Model}
\label{sec:data-model}

Our data model is based on conversion and impression events collected by user devices and grouped by the time epoch in which they occurred.  We view the data available to queriers as a database of such device-epoch groups of events, coming from many devices and defined formally as follows.

\heading{Conversion and impression events ($\mathbf{F}$).}
Consider a domain of impression events $\cI$ and a domain of conversion events $\cC$.
A set of impression and conversion events $F$ is a subset of $\cI \cup \cC$. The powerset of events is $\cP(\cI \cup \cC) := \{F: F \subset \cI \cup \cC\}$.

\heading{Device-epoch record ($\mathbf{x}$).}
Consider a set of epochs $\cE$ and a set of devices $\cD$.
We define the domain for device-epoch records $\cX := \cD \times \cE \times \cP(\cI \cup \cC)$.
That is, a {\em device-epoch record} $x = (d,e,F)$ contains a device identifier $d$, an epoch identifier $e$, and a set of impression and conversion events $F$.

\heading{Database ($\mathbf{D}$).}
A {\em database} is a set of device-epoch records, $D \subset \cX$, where a device-epoch appears at most once.  That is, $\forall d,e \in \cD \times \cE, |\{F \subset \cI \cup \cC: (d, e, F) \in D\}| \le 1$.  We denote the set of all possible databases by $\D$.  This will be the domain of queries in \sysname.
Given a database $D \in \D$ and $x \in \cX$, $D + x$ denotes that device-epoch record $x$ is added to database $D$ that initially did not include it.

\heading{Device-epoch events data ($\mathbf{D_d^e}$, $\mathbf{D_d^E}$).}
Given a database $D \in \D$, we define $D_d^e \subset \cI \cup \cC$ as $D_d^e = F$ if there exist (a unique) $F$ such that $(d,e,F) \in D$, and $D_d^e = \emptyset$ otherwise.  Think of this as the event data of device $d$ at epoch $e$. 
We also define $D_d^E := (D_d^e)_{e \in E} \in  \cP(\cI \cup \cC)^{|E|}$
the events of device $d$ over a set of epochs $E$ (typically a contiguous window of epochs).

\heading{Public events ($P$).}
A key innovation in \sysname's data model is to support incorporation of side information that can be reliably assumed as available to the querier. %
For example, an advertiser such as Nike can reliably know when someone places a product into a cart (i.e, a conversion occurred), though depending on whether the user is logged in or not, Nike may or may not know who did that conversion.

We model such side information as a domain of {\em public events} for a querier, denoted $P \subseteq \cI \cup \cC$. $P$ is a subset of all possible events, that will be disclosed to the querier if they occur in the system.  We do {\em not} assume that the querier knows the devices on which events in $P$ occur, and different queriers can have knowledge about different subsets of events.
Such side information is typically not modeled explicitly in DP systems, as DP is robust to side information. \sysname also offers such robustness to generic side information. However, we find that additionally modeling the ``public'' events known to the querier has two key benefits. First, it opens DP optimizations that leverage this known information to consume less privacy budget.
Second, it lets us formally define within-site linkability and adapt our design to provide a DP guarantee against such linkability. %

\vspace{-0.5em}\subsubsection{Query Model}
\label{sec:query-model}

In on-device systems, queries follow a specific format: first the attribution function runs locally to generate an attribution report, on a set of devices with certain conversions; then, the MPC sums the reports together and returns the result with DP noise. Formally, we define three concepts: attribution function, attribution report, and query. 

\heading{Attribution function, \aka attribution ($\mathbf{A}$).}
Fix a set of events relevant to the query $F_A \in \cP(\cI \cup \cC)$, and $k,m \in \N^*$ where $k$ is a number of epochs.
An {\em attribution function} is a function $A : \cP(\cI \cup \cC)^k \to \R^m$ that takes $k$ event sets $F_1, \dots, F_k$ from $k$ epochs and outputs an $m$-dimensional vector $A(F_1, \dots, F_k)$, such that only {\em relevant events} contribute to $A$. That is, for all $(F_1, \dots, F_k) \in \cP(\cI \cup \cC)^k$, we have:
\vspace{-0.3em}
$$A(F_1, \dots, F_k) = A(F_1 \cap F_A, \dots, F_k \cap F_A) . $$%
\heading{Attribution report, \aka report ($\mathbf{\rho}$).}
This is where the non-standard behavior of on-device budgeting systems, which deduct budget only for devices with specific conversions, becomes apparent.
Intuitively, we might consider attribution reports as the ``outputs'' of an attribution function. However, in the formal privacy analysis, we must account for the fact that only certain devices self-select to run the attribution function (and thus deduct budget). We model this in two steps. First, we introduce a conceptual {\em report identifier}, $r$, a unique random number that the device producing this report generates and shares with the querier at report time.

Second, we define an {\em attribution report} as a function over the whole database $D$, that returns the result of an attribution function $A$ for a set of epochs $E$ {\em only for one specific device $d$ as uniquely identified by a report identifier $r$}.  Formally, $\rho_r: D\in \D \mapsto A(D^{E}_{d})$. 
At query time, the querier selects the report identifiers it wants to include in the query (such as those associated with a type of conversion the querier wants to measure), and devices {\em self-select} whether to deduct budget based on whether they recognize themselves as the generator of any selected report identifiers. Defining attribution reports on $D$ lets us account for this self-selection in the analysis.

\heading{Query ($\mathbf{Q}$).}
Consider a set of report identifiers $R \subset \Z$, and a set of attribution reports $(\rho_r)_{r\in R}$ each with output in $\R^m$.
The {\em query} for $(\rho_r)_{r\in R}$ is the function $Q: \D \to \R^m$ is defined as $Q(D) := \sum_{r \in R} \rho_r(D)$ for $D \in \D$.

\subsubsection{Instantiation in Example Scenario}
\label{sec:data-query-model-instantiations}

To make our data and query models concrete, we instantiate the scenarios from \S\ref{sec:running-example}.

\heading{User} Ann's data, together with that of other users, populates dataset $D$. Each device Ann owns has an identifier $d$, and events logged from epoch $e$ go into observation $x = (d, e, F)$. $F = I \cup C$ is the set of all events logged on that device during that epoch, including impressions ($I$) shown to Ann by various publishers, and conversions ($C$) with various advertisers.
Other devices of Ann, other epochs, and other users' device-epochs, constitute other records in the database.

\heading{The advertiser}, Nike, can observe some of Ann's behavior on its site. As a result, any such behavior logged in $C$ on nike.com constitutes public information for querier Nike. 
This might include purchases, putting an item in the basket, as well as associated user demographics (\eg when Ann is logged-in).  However, Nike cannot observe impression or conversion events on other websites. As a result, for this querier $P = \cC_{\textrm{Nike}}$, which denotes all possible events that can be logged on nike.com. Each actual event in this set (e.g., $F \cap \cC_{\textrm{Nike}}$, including Ann's purchase) is associated with an identifier $r$ in \sysname. Using these identifiers, Nike can analyze the relative effectiveness of two ad campaigns $a_1$ and $a_2$ on a given demographics for a product $p$, such as the shoes Ann bought. First, Nike defines the set of relevant events for the shoe-buying conversion; these are any impressions of $a_1$ and $a_2$. Nike uses these relevant events in an attribution function $A : \cP(\cI \cup \cC)^{|E|} \to \R^2$ that looks at epochs in $E$ and returns, for example, the count (or value) of impression events corresponding to ads $a_1$ and $a_2$. Third, using the set of report identifiers $r$ from purchases of $p$ from users in the target demographic, Nike constructs a query $Q$ that will let it directly compare the proportion of purchases associated with ad campaign $a_1$ versus campaign $a_2$.

%% file: sections/04.2-idp-formulation.tex
\subsection{IDP Formulation and Guarantees}
\label{sec:idp-formulation-and-guarantees}

With \sysname's data and query models defined, we now formalize and prove its privacy guarantees using individual DP. After introducing our neighboring relation in \S\ref{sec:neighboring-databases}, we briefly define traditional DP for reference in \S\ref{sec:traditional-dp}, followed by individual DP in \S\ref{sec:individual-dp}. In \S\ref{sec:privacy-claims}, we state the IDP guarantees for \sysname, which imply protection against both cross-site tracking and within-site linkability.

\vspace{-0.5em}\subsubsection{Neighboring Databases}
\label{sec:neighboring-databases}

A DP guarantee establishes the neighboring database relation, determining the unit of protection. In our case, this unit is the device-epoch record. To account for the existence of public event data (\S\ref{sec:data-model}), we constrain neighboring databases to differ by one device-epoch record \emph{while preserving public information}. This ensures that a database containing an arbitrary device-epoch record is indistinguishable from a database containing a device-epoch record with the same public information but no additional data.

\heading{Neighboring databases under public information ($D \sim_x^{P} D'$).}
Given $D,D' \in \D$, $x = (e,d,F) \in \cX$ and $P \subset \cI \cup \cC$, we write $D \sim_x^{P} D'$ if there exists $D_0 \in \D$ such that $\{D,D'\} = \{D_0 + (e,d,F), D_0 + (e,d,F \cap P)\}$.  This definition corresponds to a replace-with-default definition~\cite{FZ21} combined with Label DP~\cite{GGK+21}.  Although public data is baked into our neighboring relation, which makes it specific to each individual querier, we have proven that composition across queriers is still possible, which is important to reason about collusion (Appendix~\S\ref{sec:collusion}).

\vspace{-0.5em}\subsubsection{DP Formulation (for Reference)}
\label{sec:traditional-dp}

In DP, noise must be applied to query results based on the query's {\em sensitivity}--the worst-case difference between two neighboring databases. Traditional DP mechanisms rely on global sensitivity.

\heading{Global sensitivity.}
Fix a query $q: \D \to \R^m$ for some $m$ (so $q$ could be either a query or an individual report in our formulation).
	We define the {\em global $L_1$ sensitivity of $q$} as follows:
	\vspace{-1em}\begin{align}
		\Delta(q) := \max_{D,D' \in \D: \exists x \in \cX, D' = D + x } \|q(D) - q(D')\|_1 .
  \end{align}\vspace{-0.5em}

\heading{Device-epoch DP.}
When scaling DP noise to the global sensitivity under our neighboring definition, we can provide device-epoch DP.  Fix $\epsilon>0$ and $P \subset \cI \cup \cC$. 
A randomized computation $\cM : \D \to \R^m$ satisfies {\em device-epoch $\epsilon$-DP} if for all databases $D, D' \in \D$ such that $D \sim_x^{P} D'$ for some $x \in \cX$, for any set of outputs $S \subseteq \R^m$ we have $\Pr[\cM(D) \in S] \le e^\epsilon \Pr[\cM(D')\in S]$.  This is the traditional DP definition, instantiated for our neighboring relation.  %

\vspace{-0.5em}\subsubsection{IDP Formulation}
\label{sec:individual-dp}

Since queries are aggregated from reports computed on-device with known data, we would prefer to scale the DP noise to the individual sensitivity, which is the worst case change in a query result triggered by the specific data for which we are computing a report.

\heading{Individual sensitivity.}
Fix a function $q: \D \to \R^m$ for some $m$ (so $q$ could be either a query or an individual report in our formulation) and $P \subset \cI \cup \cC$.
Fix $x \in \cX$.
	We define the {\em individual $L^1$ sensitivity of $q$ for $x$} as follows:
\vspace{-0.5em}	\begin{align}
		\Delta_x(q) := \max_{D,D' \in \D: D' = D + x } \|q(D) - q(D')\|_1 .
	\end{align}\vspace{-1em}

While we cannot directly scale the noise to individual sensitivity, we can scale the on-device budget consumption using this notion of sensitivity.
That is, for a fixed and known amount of noise that will be added to the query, a lower individual sensitivity means that less budget is consumed from a device-epoch. This approach provides a guarantee of individual
\footnote{
While referred to as Personalized Differential Privacy (PDP) in some papers  \cite{ESS15}, we use the term Individual Differential Privacy (IDP), as it better reflects the concept and aligns with individual sensitivity, the basis of the definition. 
This recent paper \cite{FZ21} also uses IDP terminology.
}
DP \cite{ESS15,FZ21} for a device-epoch, defined as follows.

\heading{Individual device-epoch DP.}
Fix $\epsilon>0$,  $P \subset \cI \cup \cC$, and $x \in \cX$.
A randomized computation $\cM : \D \to \R^m$ satisfies {\em individual device-epoch $\epsilon$-DP for $x$} if for all databases $D, D' \in \D$ such that $D \sim_x^{P} D'$, for any set of outputs $S \subseteq \R^m$ we have $\Pr[\cM(D) \in S] \le e^\epsilon \Pr[\cM(D')\in S]$.

Intuitively, IDP ensures that, from the point of view of a fixed device-epoch $x$, the associated data $F$ is as hard to recover from query results as it would be under DP. %

\vspace{-0.5em}\subsubsection{IDP Guarantees}
\label{sec:privacy-claims}
Through IDP, we prove two main properties of \sysname: (1) {\bf Individual DP guarantee}, which implies bounds on {\em cross-site leakage}, demonstrating that the API cannot be used to reveal cross-site activity; and (2) {\bf Unlinkability guarantee}, which implies bounds on {\em within-site linkability}, demonstrating that the API cannot be used even by a first-party site to distinguish whether a set of events is all on one device vs. spread across two devices.
Proofs are in Appendix~\S\ref{sec:appendix:privacy-guarantees-proofs}.

For the IDP guarantee, we give two versions. First, a stronger version under a mild constraint on the class of allowed queries, specifically that $\forall i, \forall F, \ A(F_1,\mydots, \allowbreak F_{i-1},F_i \cap P,\allowbreak F_{i+1}, \mydots, F_k) = A(F_1,\mydots,F_{i-1},\emptyset,F_i, \mydots, F_k)$. A sufficient condition is to ensure that queries leverage public events only through their report identifier, \ie $F_A \cap P = \emptyset$. The queries from the scenarios we consider (\S\ref{sec:running-example}) satisfy this property.
Second, a slightly weaker version of the DP guarantee with increased privacy loss, but with no constraints on the query class, which is useful when considering colluding queriers.

\vspace{-0.3em}\begin{theorem}[{\bf Individual DP guarantee}]
	\label{thm:e2e_idp}
        Fix a set of public events $P \subset \cI \cup \cC$, and
        budget capacities $(\epsilon^G_d)_{d \in \cD}$. 
        {\bf Case 1:}
        If all the queries use attribution functions $A$ satisfying $\forall i, \forall F, \ A(F_1,\mydots,F_{i-1},F_i \cap P,F_{i+1}, \mydots, F_k) = A(F_1,\mydots,F_{i-1},\emptyset,F_i,\newline \mydots, F_k)$, then for $x \in \cX$ on device $d$, \sysname satisfies individual device-epoch $\epsilon^G_d$-DP for $x$ under public information $P$.
        {\bf Case 2:}
        For general attribution functions, \sysname satisfies individual device-epoch $2\epsilon^G_d$-DP for $x$ under public information $P$.
\end{theorem}

\vspace{-0.5em}Intuitively, the information gained on cross-site (private to the querier) events in device-epoch $x$ under the querier's queries is bounded by $\epsilon^G_x$ (or $2\epsilon^G_x$ without query constraints).

\vspace{-0.5em}\begin{theorem}[{\bf Unlinkability guarantee}]
    \label{thm:unlinkability}
        Fix budget capacities $(\epsilon^G_d)_{d \in \cD}$.
        Take any $d_0, d_1 \in \cD$, $e \in \cE$, and $F_1 \subset F_0$.
        Denote $x_0 := (d_0, e, F_0), x_1 := (d_1, e, F_1), x_2 := (d_0, e, F_0 \setminus F_1) \in \cX$.
        For any $D,D' \in \D$ such that $\{D,D'\} = \{D_0 + x_0, D_0 + x_1 + x_2\}$ for some $D_0 \in \D$,
        instantiation $\cM$ of \sysname, %
        and $S \subset Range(\cM)$ we have:
           $\Pr[\cM(D) \in S] \le e^{2\epsilon^G_{d_0} + \epsilon^G_{d_1}} \Pr[\cM(D') \in S] .$
        
\end{theorem}

\vspace{-0.5em}Intuitively, linking a set of events across two devices---compared to detecting these events on one device---is only made easier by the amount of budget on the second device; \sysname does not introduce additional privacy loss for linkability, above what is revealed through DP queries.

\vspace{-0.3em}

%% file: sections/04.3-idp-optimizations.tex
\subsection{IDP Optimizations}
\label{sec:idp-optimizations}

IDP allows discounting the DP budget based on individual sensitivity, which is never greater but often smaller than global sensitivity. The easiest way to grasp this opportunity is to visualize and compare the definitions of global and individual sensitivities for reports and queries. Recall that \sysname enforces a bound on reports by capping each coordinate in the attribution function's output to a querier-provided maximum. Given this cap, we prove the following formulas for both sensitivities (proofs in Appendix~\S\ref{sec:appendix:idp-optimizations-proofs}):

\vspace{-0.5em}\begin{theorem}[{\bf Global sensitivity of reports and queries}]
	\label{thm:global_sensitivity_of_reports_queries}
	Fix a report identifier $r$, a device $d_r$, a set of epochs $E_r$, an attribution function $A$ and the corresponding report $\rho: D \mapsto A(D^{E_r}_{d_r})$.
	We have:
	\begin{align*}
		\Delta (\rho) = \underset{i \in [k], F_1, \dots, F_k \in \cP(\cI \cup \cC)}{\max \|A(F_1, \mydots, F_k) } - A(F_1, \mydots, F_{i-1}, \emptyset, F_{i+1}, \mydots, F_k) \|_1
	\end{align*}

        Next, fix a query $Q$ with reports $(\rho_r)_{r\in R}$ such that each device-epoch participates in at most one report.
        We have  $\Delta(Q) = \max_{r \in R} \Delta(\rho_r)$.
        
\end{theorem}

\vspace{-0.6em}\begin{theorem}[{\bf Individual sensitivity of reports and queries}]
	\label{thm:individual_sensitivity_of_reports_queries}
	Fix a device-epoch record $x = (d,e,F) \in \cX$.
	Fix a report identifier $r$, a device $d_r$, a set of epochs $E_r = \{e_1, \dots, e_k\}$, an attribution function $A$ with relevant events $F_A$, and the corresponding report $\rho: D \mapsto A(D^{E_r}_{d_r})$.

         We have: $\Delta_x(\rho) =
			 \underset{F_1, \mydots, F_{i-1}, F_{i+1}, \mydots, F_k \in \cP(\cI \cup \cC)}{\max  \|A(F_1, \mydots, F_{i-1}, F,} F_{i+1}, \mydots, F_k) - A(F_1, \mydots, F_{i-1}, \emptyset, F_{i+1}, \mydots, F_k) \|_1  $ if $ d = d_r$ and $e = e_i \in E_r$, and 			$\Delta_x(\rho) = 0$                 otherwise.

        In particular,
 	\begin{align*}
		\Delta_x(\rho) \le
		\begin{cases}
                0     & \text{if } d = d_r, e \in E_r \text{ and } F \cap F_A = \emptyset \\
                \|A(F) - A(\emptyset)\|_1 & \text{if } d = d_r \text{ and } E_r = \{e\} \\
                \Delta(\rho) & \text{if } d = d_r, e \in E_r \text{ and } F \cap F_A \neq \emptyset 
		\end{cases}
	\end{align*}

        Next, fix a query $Q$ with reports $(\rho_r)_{r\in R}$.  Then we have:
            $\Delta_x(Q) \le \sum_{r \in R} \Delta_x(\rho_r)$.
        In particular, if $x$ participates in at most one report $\rho_r$, then:
            $\Delta_x(Q) = \Delta_x(\rho_r)$.
\end{theorem}

\vspace{-0.5em}
This theorem justifies both the inherent optimization used by all on-device systems and the new optimizations added in \sysname.

\heading{Inherent on-device optimization.}
The condition $d = d_r$ in \Thm\ref{thm:individual_sensitivity_of_reports_queries} explains why, under IDP, on-device budgeting systems deduct privacy loss only for devices that participate in a query. This is more efficient than off-device systems like \IPA, which, under traditional DP, must deduct budget based on $\Delta(Q)$ from {\em all devices}, regardless of their participation (\Thm\ref{thm:global_sensitivity_of_reports_queries}).

\heading{New optimization examples.}
First, devices that participate in a query but have no relevant data (\ie $F \cap F_A = \emptyset$ or $A(F) = A(\emptyset)$ in \Thm\ref{thm:individual_sensitivity_of_reports_queries}) do not incur budget loss. This is why, in the example from \S~\ref{sec:execution-example}, we don't deduct from epoch $e_3$, which has no Nike impressions. 
Second, a device's individual sensitivity depends only on reports it participates in ($\Delta_x(Q) = \Delta_x(\rho_r)$), whereas global sensitivity depends on all reports in the query ($\Delta(Q) = \max_{r \in R} \Delta(\rho_r)$). For instance, since the report $\rho$ typically depends on the public information $F \cap P$ of a record $(d,e,F)$, we use a \$70 cap instead of \$100 in the Nike example.
Third, if an attribution spans only one epoch (or is broken into single-epoch reports), individual sensitivity can be further reduced based on the private information $F$. For example, if Nike measures the average impression-to-conversion delay (0 to 7 days) in a single epoch and a record $x$ has one impression only 1 day before the conversion, its individual budget will be 1/7th of the global budget.

%% file: sections/05-prototype.tex
\vspace{-0.6em}\section{Chrome Prototype}
\label{sec:prototype}

We integrated \sysname into Google Chrome by modifying ARA. We disabled \ARA's impression-level budgeting, added epoch support, and extended ARA's database to include a table for privacy filters for each epoch-querier pair. Unlike \ARA, which supports only last-touch attribution and fetches only the latest impression, our implementation retrieves all impressions related to the conversion, groups them by epoch, and identifies epochs with no relevant data to avoid unnecessary budget consumption.

%% file: sections/06-evaluation.tex
\vspace{-0.7em}
\section{Evaluation}
\label{sec:evaluation}

We seek to answer three key questions: 
\vspace{-0.3em}
\begin{enumerate}
	\item[{\bf Q1:}] How do optimizations impact  budget consumption?
	\item[{\bf Q2:}] How do optimizations impact query accuracy?
        \item[{\bf Q3:}] How effective is bias measurement?
\end{enumerate}

We also evaluate \sysname's runtime overhead, but provide these results in Appendix~\ref{sec:appendix:performance-and-bandwidth-overheads} as they are not vital to our main hypotheses in this paper.

\vspace{-0.5em}\subsection{Methodology}
\label{sec:evaluation:methodology}

We evaluate \sysname on three datasets—a microbenchmark and two realistic advertising datasets from PATCG and Criteo—and compare its privacy budget consumption and query accuracy against two baselines. The first baseline is {\bf \baselineIPA}, our own prototype implementing IPA's centralized budgeting and query execution. The second is {\bf \baselineARA}, a version of \ARA providing device-epoch-level guarantees. \baselineARA includes the inherent optimization of all on-device systems but excludes the new optimizations in \S\ref{sec:idp-optimizations}.

\heading{Scenario-driven methodology.}
We conduct our evaluation by enacting the scenario from \S\ref{sec:running-example}. An advertiser (Nike) runs ad campaigns and repeatedly measures their efficacy. Each time a customer purchases quantity $C$ of a product, Nike requests an attribution report, specifying the relevant ad campaigns. Nike requests reports over some attribution window and uses last-touch attribution. If no relevant impression is found, the report value is $0$; otherwise, it is $C$. Nike batches reports and submits them to the aggregation service for a DP summation query using the Laplace mechanism. In our experiments, Nike repeatedly performs queries on report batches of size $B$, which varies by dataset. Once $B$ reports are gathered, Nike runs its query. This is repeated over time as more batches of $B$ reports are gathered.  This is also repeated for each product, e.g., 10 in the microbenchmark/PATCG and a variable number in Criteo.

When requesting an attribution report for a conversion, Nike must specify the requested privacy budget, $\epsilon$ -- the same value for all reports in a batch.  Since the MPC uses the Laplace mechanism to ensure $\epsilon$-DP, Nike selects $\epsilon$ to achieve acceptable accuracy. We assume Nike chooses $\epsilon$ in an attempt to keep query error within 5\% ($\alpha = 0.05$) of the true value with 99\% probability ($\beta = 0.01$), which corresponds to roughly 0.02 RMSRE. The formula for $\epsilon$ is: $\epsilon = \Delta \ln(1/\beta) / (\alpha \cdot B \cdot \tilde{c})$, where $\Delta$ is the maximum value for $C$ and $\tilde{c}$ is Nike's rough estimate of the average $C$.

Our specific method is: we run repeated, single-advertiser summation queries on fixed-size batches of attribution reports, using last-touch attribution and a privacy budget calibrated as described above. Default parameters include: a 7-day epoch size, a 30-day attribution window, and a global privacy budget per epoch of $\epsilon_G=1$.

\heading{Microbenchmark dataset.}
To methodically evaluate \sysname, under a range of conditions, more or less favorable to our optimizations, we create a synthetic dataset with 40,000 conversions across 10 products over 120 days. We expose two knobs: {\bf Knob1}, the user participation rate per query, determines the fraction of users who are assigned conversions relevant for a particular query; {\bf Knob2}, the number of impressions per user per day. These knobs impact budget allocation across \baselineIPA, \baselineARA, and \sysname. Lower Knob1 increases opportunities for fine-grained accounting in \baselineARA and \sysname. Lower Knob2 allows \sysname to conserve privacy by not deducting from epochs with no relevant impressions, a key optimization over \baselineARA.

\heading{\patcg dataset.}
To evaluate \sysname under more realistic conditions, we resort to the PATCG and Criteo datasets.  PATCG is a synthetic dataset released by the namesake W3C community group~\cite{patcg-synthetic-dataset}, which contains 
24M conversions from a single advertiser over 30 days. This dataset represents a large advertiser, with only 1\% of conversions attributed to impressions. There are 16M distinct users, and each user sees an average of 3.2 impressions. Users who convert take part in 1.5 conversions on average.

\heading{\criteo dataset.}
The Criteo dataset~\cite{tallis2018reacting} is sampled from a 90-day log of live ad impressions and conversions recorded by the Criteo ad-tech. The dataset includes data from 292 advertisers with 12M impression records and 1.3M conversion records. There are 10M unique users.  The dataset provides opportunities for evaluating \sysname in some additional dimensions compared to \patcg and the microbenchmark.  In particular, the \criteo dataset contains data from multiple advertisers of widely distinct sizes, i.e., having a wide range in terms of number of impressions (1--2.6M impressions) and conversions per advertiser (0--478k conversions). 
However, since the dataset is heavily subsampled, missing many impressions, we also evaluate \sysname on augmented versions of this dataset, in which we add synthetic impressions to compensate for the missing impressions that might otherwise favor \sysname’s optimizations.

\begin{figure*}[t!]
\captionsetup[subfloat]{captionskip=-0.3cm}
\setlength{\abovecaptionskip}{4pt}
\centering
\includegraphics[width=\linewidth]
{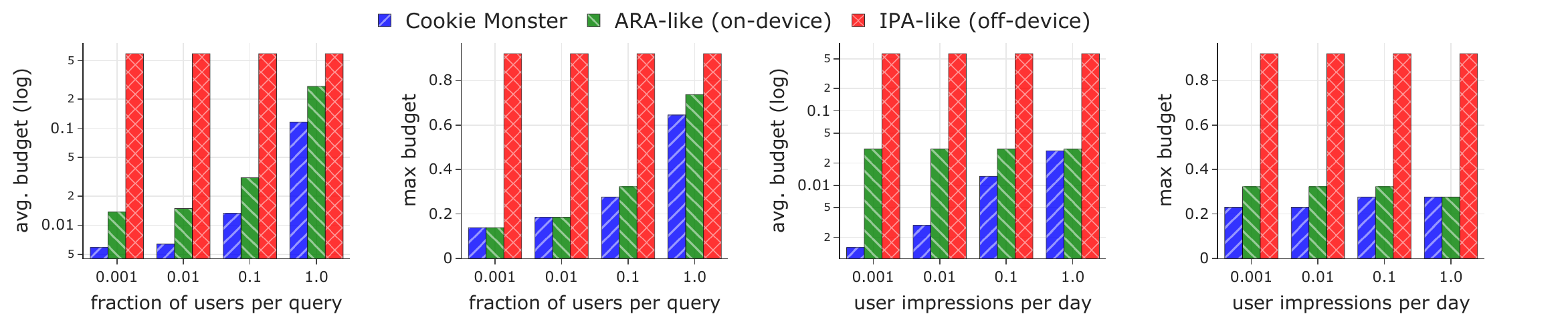}
\subfloat[Avg. budget varying knob1\label{fig:microbenchmark_avg_vary_knob1}]{\hspace{.25\linewidth}}
\subfloat[Max. budget varying knob1\label{fig:microbenchmark_max_vary_knob1}]{\hspace{.25\linewidth}}
\subfloat[Avg. budget varying knob2\label{fig:microbenchmark_avg_vary_knob2}]{\hspace{.25\linewidth}}
\subfloat[Max. budget varying knob2\label{fig:microbenchmark_max_vary_knob2}]{\hspace{.25\linewidth}}
\caption{{\bf Budget consumption on the \microbenchmark.} (a) and (b) show average and maximum budget consumption across all device-epochs, respectively, as a function of the fraction of users that participate per query (knob1); value of knob2 is constant 0.1. (c) and (d) show the same metrics as a function of user impressions per day (knob2); value of knob1 is constant 0.1.}
\label{fig:microbenchmark_knobs}
\end{figure*}

\vspace{-0.7em}\subsection{Microbenchmark Evaluation (Q1)}
\label{sec:evaluation:microbenchmark}

We use the \microbenchmark to evaluate the impact of individual-sensitivity optimizations on privacy budget consumption across a range of controlled workloads (question Q1).

\heading{Varying user participation rate per query (knob1).} We first vary the user participation rate per query. With a default batch size of 2,000 reports and 10 products (queried twice, totaling 20 queries), we create 40,000 conversions. Knob1 controls how these conversions are assigned to users, indirectly determining the total number of users. A lower knob1 favors on-device budgeting, as it spreads the 40,000 conversions across more users, creating more privacy filters for the advertiser. For example, with knob1 = 1, each user participates in all 20 query batches, requiring a minimum of 2,000 users, while knob1 = 0.001 generates 2M users. In the \patcg dataset, users convert with a 0.05 daily rate, corresponding to knob1 = 0.1, which we use as default in other experiments.

\Fig\ref{fig:microbenchmark_avg_vary_knob1} and \ref{fig:microbenchmark_max_vary_knob1} show the average and maximum budget consumption across all device-epochs requested through the 20 queries. Qualitatively, the average budget consumption is a much more useful metric to assess the efficiency of the three systems, but we include the maximum because it reduces IDP guarantees to standard DP guarantees, thereby providing a more apples-to-apples comparison between on-device and off-device budgeting. Recall that \baselineIPA does not distribute budget consumption across devices but has a centralized privacy filter for each epoch, from which it deducts budget upon executing each query.  As a result, increasing user participation per query (knob1) does not impact its budget consumption, which is always higher than the other methods'.  \sysname consistently consumes the least budget due to its optimizations, with greater improvements as user participation increases (lower knob1), since more device-epochs lack relevant impressions and don't deduct budget. Even under the max budget metric, on-device systems outperform \baselineIPA, with \sysname being the most efficient.

\heading{Varying the number of impressions per user per day (knob2).} 
We now fix knob1 at 0.1 and vary the number of impressions per user per day (knob2). In \patcg, users see an average of 3.22 ads over 30 days, giving knob2 a value of 0.1. \Fig\ref{fig:microbenchmark_avg_vary_knob2} and \ref{fig:microbenchmark_max_vary_knob2} confirm that \sysname's optimizations are most effective when users have fewer impressions.

Thus, \sysname reduces budget consumption compared to baselines, especially when budget is spread across many users and when users have fewer impressions.

\vspace{-0.5em}\subsection{\patcg Evaluation (Q1, Q2)}
\label{sec:evaluation:patcg}

\begin{figure*}[t!]
\captionsetup[subfloat]{captionskip=-0.3cm}
\setlength{\abovecaptionskip}{4pt}
\centering
\includegraphics[width=\linewidth]{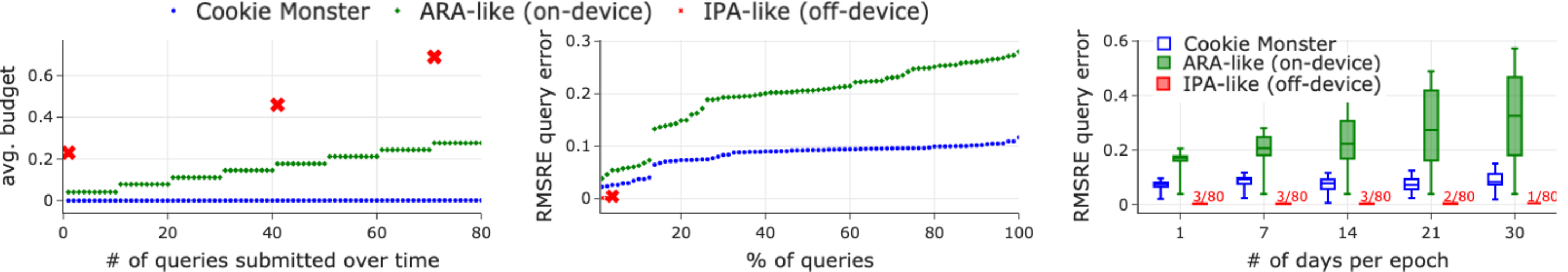}
\subfloat[Avg. budget consumed across all device-epochs\label{fig:patcg_avg_budget_consumption}]{\hspace{.33\linewidth}}
\subfloat[CDF of RMSRE\label{fig:patcg_rmsre_cdf}]{\hspace{.33\linewidth}}
\subfloat[RMSRE as a function of epoch length\label{fig:patcg_rmsre_boxes}]
{\hspace{.33\linewidth}}
\caption{{\bf Budget consumption and query accuracy on the \patcg dataset.} (a) Average budget consumption across all device-epochs as a function of the number of queries submitted by the advertiser. (b) CDF of RMSRE with a 7-day epoch. (c) RMSRE median (horizontal lines), first and third quartiles (boxes), and max/min (top/bottom range markers) as epoch length increases.}
\label{fig:patcg_experiments}
\end{figure*}

We use the \patcg dataset to evaluate \sysname's impact on budget consumption (Q1) and query accuracy (Q2). This dataset links impressions and conversions to attributes, with values uniformly sampled from 0 to 9, representing 10 potential products. Nike queries each product eight times over the four months spanning the dataset, totaling 80 queries with batch sizes between 280,000 and 303,009 reports. Large batch sizes accommodate the low attribution rate (1\% of impressions relevant to conversions), assuming Nike adjusts batch sizes accordingly.

\Fig~\ref{fig:patcg_avg_budget_consumption} illustrates the average privacy budget consumed by each system as 80 queries are submitted for execution by the advertiser. The x-axis represents the order of queries, with points indicating budget consumption. \baselineIPA executes only a small fraction of queries (3.75\%) due to its coarse-grained, population-level accounting, leading to early budget depletion. \baselineARA and \sysname, with finer-grained, individual-level accounting, execute all queries and resulting in smoother and lower average budget consumption. \sysname shows up to 206 times lower average budget consumption compared to \baselineARA, highlighting the benefits of its individual-sensitivity optimizations.

Next, we assess query accuracy (Q2). On-device systems (\baselineARA and \sysname) hide budgets when depleted, which can affect query accuracy, while \baselineIPA explicitly rejects queries with exhausted budgets. As in our experiments, privacy budgets are set to aim for high accuracy in the Laplace mechanism, we expect IPA's executed queries to have errors within the 0.02 mark. In contrast, \ARA and \sysname may incur additional errors when epochs run out of budget, leading to nullified or incomplete reports.

\Fig~\ref{fig:patcg_rmsre_cdf} shows the CDF of root mean square relative error (RMSRE), defined as $\sqrt{\E [ (\cM(D) - Q(D))^2/Q(D)^2] }$ for an estimate $\cM(D)$ of the query output $Q(D)$. This metric captures both Laplace-induced and IDP-bias-induced errors. The CDF shows query errors for each system. \baselineIPA's line ends at 3.75\% of queries, aligning with its budget constraints but maintaining within the 5\% error mark. \sysname consistently exhibits lower errors than \baselineARA due to its budget conservation, resulting in fewer nullified reports and reduced bias.  This is true without any bias mitigation strategies.  In \S\ref{sec:evaluation:bias-detection}, we show that even with bias measurement running alongside every query, \sysname still outperforms \baselineARA (which has no bias measurement) in terms of budget consumption and query accuracy.

Finally, we explore how epoch length affects performance. Longer epochs strengthen device-epoch privacy guarantees but slow budget refreshing, leading to more query rejections in IPA and increased bias in on-device systems without mitigation. \Fig~\ref{fig:patcg_rmsre_boxes} evaluates RMSRE measures (median, first and third quartiles, and range) as epoch length varies. \baselineIPA's query execution drops to 1.25\% at one-month epochs, while \sysname and \baselineARA complete all queries but with increasing errors. \sysname's budget conservation results in fewer altered or nullified reports, maintaining lower error degradation compared to \baselineARA as epochs grow.

\begin{figure*}[t!]
\captionsetup[subfloat]{captionskip=-0.3cm}
\setlength{\abovecaptionskip}{4pt}
\centering
\includegraphics[width=\linewidth]{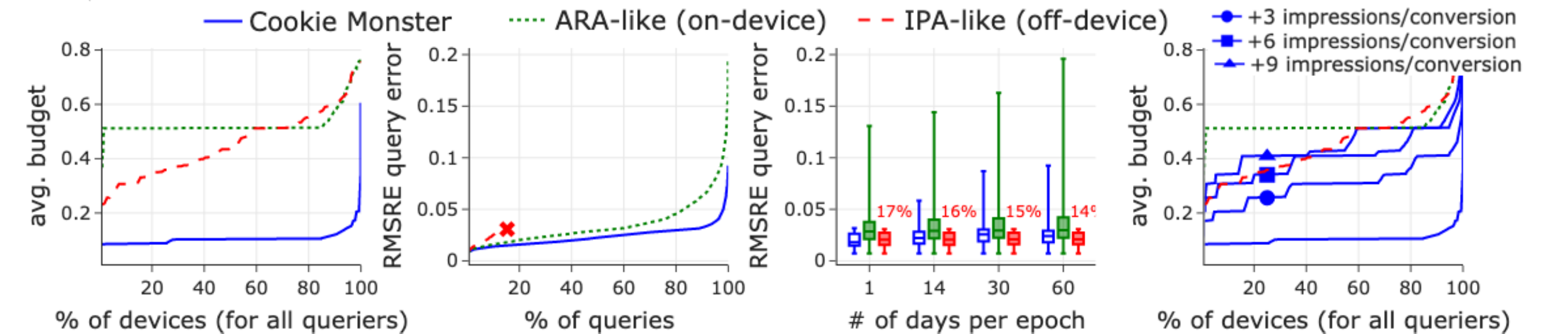}

\subfloat[CDF of budget on \criteo.]{\label{fig:criteo_experiments:budget_cdf}\hspace{.25\linewidth}}
\subfloat[CDF of RMSRE.]{\label{fig:criteo_experiments:rmsre_cdf}\hspace{.25\linewidth}}
\subfloat[RMSRE as function of epoch length.]{\label{fig:criteo_experiments:rmsre_boxes}\hspace{.25\linewidth}}
\subfloat[CDF of budget on \criteoplus.]{\label{fig:criteo_experiments:budget_cdf_criteoplus}\hspace{.25\linewidth}}
\caption{
{\bf Budget consumption and query accuracy on \criteo.} 
(a) CDF of per-device average budget consumption across epochs for all devices and advertisers. 
(b) CDF of RMSREs for a 7-day epoch. 
(c) RMSRE metrics with varying epoch length (see \Fig\ref{fig:patcg_rmsre_boxes} for format). 
(d) The same CDF as in (a), but for the \criteoplus, showing the impact of synthetic impression augmentation on \sysname's performance.
}
\label{fig:criteo_experiments}
\end{figure*}

\vspace{-0.5em}\subsection{\criteo Evaluation (Q1, Q2)}
\label{sec:evaluation:criteo}

The \criteo dataset enables evaluation across diverse advertisers. It includes 1.3M conversions from 292 advertisers, with conversions ranging from 0 to 478k per advertiser. To achieve meaningful accuracy under DP, an advertiser needs a minimum number of reports. We set this minimum to 350, allowing us to formulate at least one query for 109 advertisers. Advertisers with more than 350 conversions wait to accumulate 350 reports per batch for each query, resulting in 898 queries across these advertisers using the attribute ``product-category-3'' as a product ID.

\Fig~\ref{fig:criteo_experiments:budget_cdf} shows a CDF of per-device average budget consumption across epochs, where the distribution covers all devices and all advertisers; that is, there is a single data point corresponding to each device and advertiser pair, which indicates the average consumption across epochs within an advertiser's filters on a given device by the end of the workload. Lower values indicate better performance. \sysname conserves the most privacy budget, with 95\% of device-advertiser pairs having more capacity left compared to both baselines.

\Fig~\ref{fig:criteo_experiments:rmsre_cdf} presents the CDF of RMSREs for all 898 queries. \baselineIPA completes only a small fraction of queries but with good accuracy. \baselineARA and \sysname accept all queries, potentially at the expense of higher error; however, \sysname's error distribution remains better than \baselineARA's, with errors within \baselineIPA's range for up to 96\% of queries. This results from \sysname's optimizations that conserve budget and avoid introducing bias.

\Fig~\ref{fig:criteo_experiments:rmsre_boxes} examines how RMSRE varies with epoch length. Longer epochs increase contention on per-epoch filters. Despite this, \sysname's optimizations show substantial benefits, with minimal RMSRE increase (25\% increase from 1-day to 60-day epoch for median RMSRE). Although maximum RMSRE increases with epoch length, \sysname's performance remains superior to \baselineARA.

Recall that the Criteo dataset is heavily subsampled, so there is the possibility that missing impressions may amplify the benefit of our optimizations.  To assess \sysname's performance in scenarios with more relevant impressions, we augment the \criteo dataset with synthetic impressions for each conversion. The results, shown in \Fig~\ref{fig:criteo_experiments:budget_cdf_criteoplus}, compare the CDFs of budget consumption with varying augmentation levels. The behavior of \baselineIPA and \baselineARA remains unchanged by augmentation, as they do not optimize for missing relevant impressions. 
For \sysname, budget efficiency decreases as more synthetic impressions are added, approaching \baselineARA's performance at 9 extra impressions per conversion. The impressions are uniformly distributed across the attribution window, ensuring that most epochs have relevant impressions for most conversions, so \sysname's optimization is eliminated and its behavior follows \baselineARA's.

\vspace{-0.6em}
\subsection{Bias Measurement (Q3)}
\label{sec:evaluation:bias-detection}

\begin{figure*}[t!]
\captionsetup[subfloat]{captionskip=-0.3cm}
\setlength{\abovecaptionskip}{4pt}
\centering
\includegraphics[width=\linewidth]{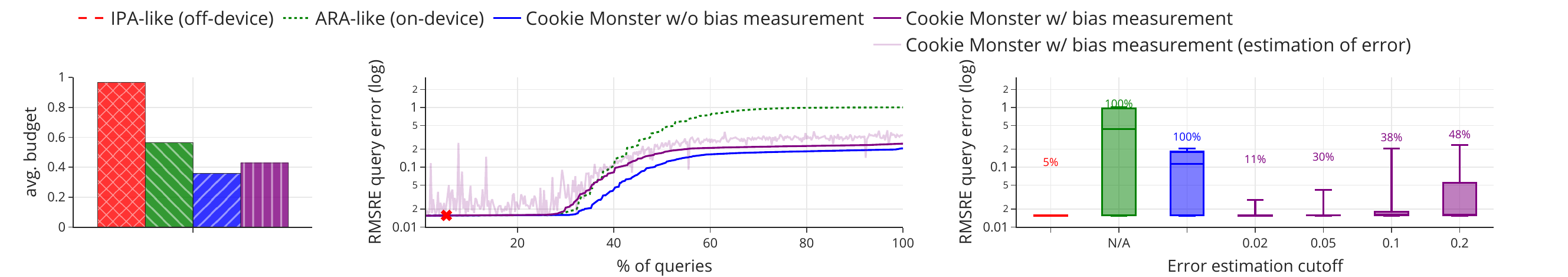}
\subfloat[Avg. budget consumed]{\label{fig:bias_experiments:budget}\hspace{.2\linewidth}}
\subfloat[CDF of RMSRE]{\label{fig:bias_experiments:budget_cdf_full}\hspace{.4\linewidth}}
\subfloat[RMSRE as function of error estimation cutoff.]{\label{fig:bias_experiments:rmsre_boxes}\hspace{.4\linewidth}}
\caption{ {\bf Budget consumption and query accuracy with bias measurement on the \microbenchmark.} (a) Average budget consumed across all device-epochs. (b) CDF of true RMSRE for executed queries, alongside \sysname's RMSRE estimation from bias measurement (light-purple line).
(c) Quartiles of true RMSRE, where queries with error estimate above a given cutoff are rejected by \sysname with bias measurement. }
\label{fig:bias_experiments}
\end{figure*}

We evaluate \sysname's bias measurement technique using our \microbenchmark with default knob settings (0.1) and an increased query load to measure significant bias. Specifically, we use 60 days and repeat each query 40 times.

\Fig\ref{fig:bias_experiments:budget} shows the budget overhead incurred by bias measurement. The bias measurement's counts are scaled to have 10\% the sensitivity of the original query, so the overall sensitivity of the query/side-query combination increases by 10\%. The average consumed budget goes from 0.36 without bias measurement to 0.43 with bias measurement; this is more than a 10\% increase since some epochs that originally paid zero budget through our IDP optimization, now pay for bias counts.

\Fig\ref{fig:bias_experiments:budget_cdf_full} shows the CDF of RMSREs across all 400 queries, with a log scale on the y-axis to highlight smaller differences among \sysname variants compared to \ARA. Due to the heavy query load, \IPA executes only 5\% of the queries and \ARA ultimately returns empty reports, resulting in a relative error of 1.   \sysname without bias measurement plateaus at 0.2 error.
\sysname with bias measurement shows a similar trend to \sysname without it, albeit with increased error, because the higher sensitivity of the query leads additional epochs to run out of budget. However, the bias measurements let queriers compute an estimate of the error, which, although noisy (as it is also differentially private), generally serves as an upper bound on true RMSRE. Queriers can compare this estimate to a predetermined cutoff and reject queries exceeding it.
\Fig\ref{fig:bias_experiments:rmsre_boxes} displays the quartiles of true RMSREs after rejecting queries based on estimated RMSRE cutoffs. For instance, using a cutoff of 0.05 enables queriers to limit bias, achieving a maximum error of 0.04 (down from 0.21), but only accepting 30\% of the queries. 
Rejected queries still consume budget, as rejection is a post-processing step.

Thus, even with rudimentary bias measurement, \sysname offers substantial benefits over IPA while maintaining lower real error than ARA.
While we validated our technique on a microbenchmark with increased query load, applying it to real-life datasets remains an open challenge.
Future work could enhance our technique by scheduling bias measurements or using DP threshold comparison mechanisms.

%% file: sections/07-related-work.tex
\vspace{-0.8em}\section{Related Work}
\label{sec:related-work}

\heading{DP systems.}
Most DP systems operate in the centralized-DP model, where a trusted curator runs queries using global sensitivity~\cite{DR13}. Some implement fine-grained accounting through parallel composition \cite{McS09,sage,privatekube,cohere}, a coarse form of individual DP (IDP) that lacks optimizations like those in \sysname. Others function in the local-DP model, where devices randomize their data locally~\cite{local_dp_2}, and therefore inherently do on-device budgeting but have higher utility costs.
Distributed systems like~\cite{orchard, arboretum} emulate the central model with cryptographic constructions; like IPA, they maintain a single privacy filter, not leveraging IDP to conserve budget.
\cite{prochlo} uses the shuffle model~\cite{shuffle_dp} to combine local randomization with a minimal trusted party. \sysname operates in the central model with on-device budgeting and uses an IDP formalization to enable new optimizations.

\heading{Private ads measurement.}
Several proposals exist for private ad measurement systems. Apple's PCM~\cite{PCMblog2021} relies on entropy limits for privacy. Meta and Mozilla's IPA~\cite{ipa-paper} uses centralized budgeting, while Google's ARA~\cite{ara} and Apple's PAM~\cite{pam} utilize on-device budgeting. ARA has primarily focused on optimizing in-query budget and utility. \cite{DGK+23} optimizes a single vector-valued hierarchical query, whereas \cite{AGK+23} assumes a simplified ARA with off-device impression-level DP guarantees, efficiently bounding each impression's contribution for  queries known upfront. \cite{delaney2024differentially} offers a framework for attribution logic and DP neighborhood relations, proposing clipping strategies for bounding global sensitivity. Our work optimizes on-device budgeting across queries, using tighter individual sensitivity bounds.  Our method is agnostic to how these bounds are enforced, potentially benefiting from clipping algorithms~\cite{DGK+23, AGK+23, delaney2024differentially}.

IDP was introduced in the centralized-DP setting, where a trusted curator manages individual budgets and leverages individual sensitivity to optimize privacy accounting \cite{ESS15,FZ21}. IDP is used for SQL-like queries and gradient descent. The literature emphasizes the need to keep individual budgets private. \cite{YKK+23} studies the release of DP aggregates over these budgets while \cite{ESS15} notes that out-of-budget records must be dropped silently, leaving bias analysis for future work.

%% file: sections/08-conclusion.tex
\vspace{-0.5em}\section{Conclusion}
\label{sec:conclusion}

Web advertising is at a crossroads, with a unique opportunity to enhance online privacy through new, privacy-preserving APIs from major browser vendors. 
We show that a novel individual DP formulation can significantly improve privacy budgeting in on-device systems. However, further progress is needed in query support, error management, and scalability. Our paper provides foundational insights and formal analysis to guide future research and industry collaboration. 

\section*{Acknowledgements}
We thank the anonymous reviewers for their constructive feedback and the Meta and Mozilla IPA teams, particularly Ben Savage and Martin Thompson, for their ongoing input. Special thanks to PATCG participants, especially Luke Winstrom, for their feedback on early proposals. This work was supported by NSF grants EEC-2133516, CNS-2106530, CNS-2104292, NSERC RGPIN-2022-04469, Google, Microsoft, Sloan Faculty Fellowships, and an Onassis Foundation Scholarship. Co-author Geambasu was partially employed by Meta during this project.

%% file: appendix/01-additional-use-cases.tex
\section{Additional Use Cases}
\label{sec:appendix:additional-use-cases}

Our scenario from \S\ref{sec:running-example} included the limited perspective of a single advertiser, Nike. Correspondingly, our system execution example (\S\ref{sec:execution-example}) and formal-model instantiation example (\S\ref{sec:data-query-model-instantiations}) focused only on Nike's perspective. However, there are many other players, with distinct perspectives, in the Web advertising ecosystem, such as: first-party content providers that are also advertising platforms, like Meta, which seek to ad placement from multiple advertisers; and third-party ad-techs like Criteo that seek to optimize ad placement across many publishers and advertisers.  In this section we discuss the Meta perspective, which our theory can readily support.  We are still working on reasoning through the theory to support an intermediary ad-tech perspective.

\heading{Ad-tech perspective.}
In addition to advertising on nytimes.com, Nike also advertises on Meta, a content provider (\aka publisher or ad-tech) that runs its own, in-house advertising platform.  Ann uses Meta's facebook.com site to read posts related to running and other interests.  To show her the most relevant ads, the site requires her to log into her account and then tracks her activity within the site to build a profile of her interests.  Ann accepts that Meta learns about her interests as she interacts with content on the site while logged into her account; however, Ann expects Meta not to be tracking her across other sites on the Web, and also to not be linking her interactions as part of different accounts.  For example, while Meta may learn that Ann is passionate about running, and hence may show her the Nike running-shoe ad, Meta should not be able to tell whether Ann later buys the shoes, as that conversion occurs on nike.com. Still, to maximize the effectiveness of ads (and return on Nike's ad spend), Meta needs to be able to train a machine learning (ML) model that can predict, given a user profile and a context, which ad coming from which advertiser would be most effective to show, in terms of maximizing the likelihood of an eventual conversion. This model-training procedure can be thought of as bringing together many attributions reports corresponding to impressions that occur on one or more publishers (facebook.com here, but also potentially instagram.com) and conversions that occur on the many advertiser sites buying ads through Meta. This type of multi-advertiser, optimization query is a second class of queries that ad-measurement APIs aim to support without exposing cross-site information and while limiting within-site linkability (to meet expectations when the user switches accounts).

\heading{Instantiation of ad-tech's perspective (in formal system model from \S\ref{sec:formal-system-model}).}
Meta is symmetric with Nike, on the display side. The querier's public information will be $P = \cI_{\textrm{Meta}}$.
In our scenario, Meta is interested in learning ML models to better target ads to its users, using conversions as a metric to optimize.
To this end, Meta can learn a logistic regression mapping public (to Meta) features from its users and attributes of ads (together denoted $X_d$ for device $d$), to conversion labels. This is possible under \sysname's queries by defining an attribution function $A$ that returns $X_d$ is there is a conversion, zero otherwise, and using algorithms to fit logistic regressions under known features but private labels \cite{patcg-ipa-walr}.

%% file: appendix/o6-performance-evaluation.tex
\section{\sysname Performance Overhead}
\label{sec:appendix:performance-and-bandwidth-overheads}

We measure the performance overhead of \sysname compared to Google's \ARA. \sysname iterates over all impressions relevant to a conversion to identify which privacy filters will consume budget, whereas \ARA tracks only the most recent impression. We compare two versions of Chrome running \sysname and \ARA, using Selenium~\cite{selenium} to interact with a publisher and generate impressions across 20 epochs. Varying the number of impressions from 10 to 100, we measure the time to create a report upon triggering a conversion. \ARA consistently reports at 5.4 ms, while \sysname's reporting time increases linearly from 9.1 ms to 57.3 ms based on the number of impressions. This presents a side channel that should be addressed in the future, such as enforcing a constant runtime, to avoid revealing whether relevant impressions were found on the device.

%% file: appendix/02-system-algorithm.tex
\section{Formal Model of \sysname Algorithm}
\label{sec:appendix:e2e_algorithm}

\begin{algorithm}[ht!]
	\caption{\sysname Algorithm}
	\label{alg:e2e_functional_view}
	\begin{algorithmic}

            \Config
            \State Public events $P \subset \cI \cup \cC$
            \State Parametrized noise distribution $\cL$
		\State Device-epoch budget capacity $(\epsilon^G_x)_{x \in \cX}$
            \EndConfig

		\Input
		\State Database $D$
		\State Stream of interactively chosen queries $Q_1, \dots, Q_k$
		\EndInput

            \Function{$\operatorname{Main}$}{$D, Q_1, \dots, Q_k$}
            \State $S = \emptyset$
		\For{$(d,e,F) \in D$}
		\For{$f \in F \cap P $}
		\State Generate report identifier $r \overset{{\scriptscriptstyle\$}}{\leftarrow} U(\Z)$
		\State Save mapping from $r$ to the device that generated it: $d_r \gets d$
		\State $S \gets S \cup \{(r,f)\}$
		\EndFor
            \EndFor
		\State {\bf output} $S$ \graycomment{report identifiers and public events $D \cap P$}
            \For{$i \in [k]$}
            \State {\bf output} $\operatorname{AnswerQuery}(Q_i)$
            \EndFor
            \EndFunction
            
		\State \graycomment{Collect, aggregate and noise reports to answer $Q_i$}
            \Function{$\operatorname{AnswerQuery}$}{report identifiers $R$, target epochs $(E_{r})_{r \in R}$, attribution functions $(A_r)_{r \in R}$ and noise parameter $\sigma$}
		\For{$r \in R$}
            \State $\rho_r \gets \operatorname{GenerateReport}(d_r, E_r, A_r)$
		\EndFor
		\State Sample $X \sim \cL(\sigma)$
		\State \Return $\sum_{r \in R} \rho_r + X$
            \EndFunction

		\State \graycomment{Generate report and update on-device budget}
            \Function{$\operatorname{GenerateReport}$}{$d, E, A$}
            \For{$e \in E$}
		\State $x \gets (d,e,D_d^e)$
            \If{$\cF_x$ is not defined}
		\State Initialize filter $\cF_x$ with capacity $\epsilon^G_x$
            \EndIf
		\State $\epsilon_x \gets \operatorname{ComputeIndividualBudget}(x,d,E,A,\cL, \sigma)$
		\If{$\cF_x.\operatorname{tryConsume}(\epsilon_x) = Halt$}
		\State $F_e \gets \emptyset$
		\Else
		\State $F_e \gets D_d^e$
		\EndIf
		\EndFor
		\State $\rho \gets A((F_e)_{e \in E})$ \graycomment{Clipped attribution report}
            \State\Return $\rho$
            \EndFunction
	\end{algorithmic}
\end{algorithm}

\Alg\ref{alg:e2e_functional_view} describes the formal view of \sysname, whose privacy guarantees we establish in \S\ref{sec:privacy-claims}. \sysname answers a stream of the querier's queries by generating reports based on a device's data in the queried epochs and an attribution function $A$ passed in the query.  It does so while the querier still has available budget. The function $\operatorname{GenerateReport}$ in \Alg\ref{alg:e2e_functional_view} models this logic of privacy budget checks and consumption, followed by report creation if enough budget is available. The attribution function $A$ has bounded sensitivity (defined in \S\ref{sec:individual-dp}), enforced through clipping. Function $\operatorname{AnswerQuery}$ then sums reports together to compute the final query value. DP noise is added to the result before returning it to the querier (see the output of \Alg\ref{alg:e2e_functional_view}).

The algorithm captures the fact that reports that do not contribute to a query are not actually generated (the summation is over $r \in R$). This is how all on-device systems inherently work (not only \sysname), and it's an important optimization that preserves privacy budget, as reports that are not generated do not consume budget.  Yet, as previously mentioned, it is very non-standard behavior for DP, so its privacy justification, which we do in the next section, requires both the formalization of reports with unique identifiers $r$ and an individual DP framework.

We instantiate the filter methods and the ComputeIndividualBudget function for the Laplace distribution in the next section (\S\ref{sec:appendix:privacy-guarantees-proofs}).

%% file: appendix/03-privacy-guarantees-proofs.tex
\section{Proofs of Privacy Guarantees (\S\ref{sec:privacy-claims})}
\label{sec:appendix:privacy-guarantees-proofs}

\heading{Filter and budget semantics for Laplace.}
In this section, we focus on the Laplace noise distribution: $\cL(\sigma) = \lap(\sigma/\sqrt{2})$.
We use pure differential privacy accounting, hence the budgets are real numbers $\epsilon >0$. 
To track the budget of adaptively chosen queries, we use a Pure DP filter \cite{filters_rogers}.
For a budget capacity $\epsilon^G$, this filter simply adds up the budget consumed by the first $k$ queries, and outputs Halt for the next query with budget $\epsilon_{k+1}$ if:
\begin{align}
    \epsilon_1 + \dots + \epsilon_{k} + \epsilon_{k+1} > \epsilon^G
\end{align}

Finally, for a datapoint $x$, a report $\rho = (d,E,A)$, the Laplace distribution $\cL$ and a standard deviation $\sigma$, we have:
\begin{align}
    \operatorname{ComputeIndividualBudget}(x,d,E,A,\cL, \sigma) = \frac{\Delta\sqrt{2}}{\sigma}
\end{align}

where $\Delta$ is an upper bound on the individual sensitivity of the report $\Delta_x(\rho)$.
We provide such upper bounds in \S\ref{sec:idp-optimizations}.

Finally, we use a slightly more general way of initializing budget capacities, by setting one capacity for each possible record $(\epsilon^G_x)_{x \in \cX}$.
In the body of the paper we set the same capacity for all the records belonging to the same device $d$: $(\epsilon^G_x)_{d \in \cD}$.
For practical purposes it is enough to set capacities at the device level, but using per-record capacities simplifies certain proofs, such as \Thm\ref{thm:simplified_replace_with_public_dummy}.

\subsection{Individual DP Guarantees (\Thm\ref{thm:e2e_idp})}

\begin{algorithm}[h!]
	\caption{Inner Privacy Game}
	\label{alg:privacy_game}
	\label{alg:inner_privacy_game}

	\begin{algorithmic}

            \Config
            \State Parametrized noise distribution $\cL$
		\State Device-epoch budget capacity $(\epsilon^G_x)_{x \in \cX}$
  		\State Upper bound on number of epochs $e_{\max}$
		\State Upper bound on number of queries per epoch $k_{\max}$
            \EndConfig
 
		\Input

		\State Challenge bit $b \in \{0,1\}$
            \State Opt-out device $x_0 = (d_0, e_0, F_0) \in \cX$
		\State Adversary $\cA$

		\EndInput

		\Output
		\State View $V^b = (v_{1,1}^b, \dots, v_{1,k_{\max}}^b, v_{2,1}^b, \dots)$ of $\cA$
		\EndOutput
            \vspace{-.3cm}
            \State\hrulefill

		\State $D \gets \emptyset$
		\For{$e \in [e_{\max}]$}
		\State \graycomment{Generate data for the epoch $e$}
		\State Receive a database $G$ for epoch $e$ from $\cA$
		\If{$e = e_0$ and $(d_0, e_0) \not\in G$}
		\State $G^0 \gets G+ (d_0, e_0, \emptyset) , G^1 \gets G + (d_0, e_0, F_0)$

		\Else{}
		\State $G^b \gets G$
		\EndIf
		\State $D \gets D + G^b$

		\State
		\State \graycomment{Answer queries after epoch $e$}
		\For{$k \in [k_{\max}]$}
		\State Receive query $Q_k$ from $\cA$ with corresponding indices $R$, devices $(d_r)_{r \in R}$, target epochs $(E_{r})_{r \in R}$, attribution functions $(A_r)_{r \in R}$ and noise std-dev $\sigma$.
		\For{$r \in R$}
		\State \graycomment{Compute report for $r$}
		\For{$e \in E_r$}
		\State $x \gets (d_r,e,D_{d_r}^e)$
            \If{$\cF_x$ is not defined}
		\State Initialize filter $\cF_x$ with capacity $\epsilon^G_x$
            \EndIf
		\State $\epsilon_x \gets \operatorname{ComputeIndividualBudget}(x, d,E,A,\cL, \sigma)$
		\If{$\cF_x.\operatorname{tryConsume}(\epsilon_x) = Halt$}
		\State $F_e \gets \emptyset$
		\Else
		\State $F_e \gets D_d^e$
		\EndIf
		\EndFor
		\State $\rho_r \gets A((F_e)_{e \in E})$
		\EndFor

		\State \graycomment{Aggregate and noise reports to answer $Q_k$}
		\State Sample $X \sim \cL(\sigma)$
		\State Send $v_{e,k}^b = \sum_{r \in R} \rho_r + X$ to $\cA$

		\EndFor
		\EndFor
	\end{algorithmic}
\end{algorithm}

To prove \Thm\ref{thm:e2e_idp} from \S\ref{sec:privacy-claims}, we need to define an intermediary ``inner" privacy game \Alg\ref{alg:inner_privacy_game}, which we analyze in \Thm\ref{thm:inner_privacy}.
Next, we define another ``outer" privacy game \Alg\ref{alg:outer_privacy_game}, that is a generalized version of \Alg\ref{alg:e2e_functional_view} and internally calls \Alg\ref{alg:inner_privacy_game}.
Finally, \Thm\ref{thm:general_replace_guarantee} and \Thm\ref{thm:simplified_replace_with_public_dummy} imply \Thm\ref{thm:e2e_idp}.

\begin{theorem}[IDP of \Alg\ref{alg:inner_privacy_game} when removing $x$]
    \label{thm:inner_privacy}
    Fix a device-epoch budget capacity $(\epsilon^G_x)_{x \in \cX}$ for every possible record $x \in \cX$.
    For any opt-out record $x \in \cX$, for any adversary $\cA$, and $V^0,V^1$ defined by \Alg\ref{alg:privacy_game},
    for all $v \in \operatorname{Supp}(V)$ we have:
    \begin{align}
        \left|\ln\left( \frac{\Pr[V^0 = v]}{\Pr[V^1 = v]} \right)\right| \le \epsilon^G_x     
    \end{align}

\end{theorem}

\begin{proof}
Fix an upper bound on the number of epochs and queries per epoch $e_{\max}, k_{\max}$.
Fix an opt-out record \newline  $x = (d_0, e_0, F_0) \in \cX$ and an adversary $\cA$. Take $V^0,V^1$ the view of $\cA$ in \Alg\ref{alg:privacy_game}.
Consider a view $v \in \operatorname{Supp}(V^1)$.
We have:
    \begin{align}
        \label{eq:gen1}
        \ln \left(\frac{\Pr[V^0 = v]}{\Pr[V^1 = v]} \right) & = \ln \left( \prod_{e=1}^{e_{\max}}\prod_{k=1}^{k_{\max}} \frac{\Pr[V_{e,k}^0 = v_{e,k} | v_{<e,k}]}{\Pr[V_{e,k}^1 = v_{e,k} | v_{<e,k}]} \right)
    \end{align}

where, for $e \in [e_{\max}], k \in [k_{\max}]$, $b \in \{0,1\}$ and $v_{e,k}$ we have:

\begin{align*}
    \Pr &[V_{e,k}^b = v_{e,k} | v_{<e,k}] \\
    &= \Pr[ V_{e,k}^b  = v_{e,k} | V_{1,1}^b=v_{1,1}, \dots, V_{e,k-1}^b=v_{e,k-1}]
\end{align*}

Even though data and query parameters are adaptively chosen, they only depend on the adversary $\cA$ (fixed) and its previous views, which are fixed once we condition on $v_{<e,k}$.
Take the database ${}^{b}D^{\le e}$ and the query parameters $R$, $(\rho_r, d_r, E_r, A_r)_{r \in R}$, $\sigma$  corresponding to $\cA$ conditioned on $v_{<e,k}$. 
Note $\epsilon_{x_0}$ the state (accumulated privacy loss) of $\cF_{x_0}$ in the world with $b=1$ before answering query $e,k$.

On one hand, if $(d_0, e_0) \not \in \{(d_r,e), r \in R, e \in E_r\}$, we observe that for all $r \in R$, ${}^{0}D_{d_r}^{e_r} = {}^{1}D_{d_r}^{e_r}$, because ${}^{0}D^{\le e}$ and ${}^{1}D^{\le e}$ differ at most on $x_0 = (d_0, e_0, F_0)$.
In this case, $\forall r \in R, \rho_r({}^{0}D^{\le e}) = \rho_r({}^{1}D^{\le e})$, and hence $\Pr [V_{e,k}^0 = v_{e,k} | v_{<e,k}] = \Pr [V_{e,k}^1 = v_{e,k} | v_{<e,k}]$.

On the other hand, suppose that we have $r_1, \dots, r_\ell$ (processed in this order) such that for all $i \in [\ell]$ we have $d_{r_i} = d_0, e_0 \in E_{r_i}$. 
    
We pose $\hat R \subset R$ the set of reports that do not pass the filter in the world with $b=1$. 
(In the world with $b=0$, the filter for $(d_0, e_0, \emptyset)$ has no effect on $\rho_r({}^{0}D^{\le e})$ because whether it halts or not we have $F_{e_0} = \emptyset$).
For $r \not\in \hat R $, we have $\rho_r({}^{0}D^{\le e}) = \rho_r({}^{1}D^{\le e})$ because both worlds use $F_{e_0} = \emptyset$.

Hence, we have:

\begin{align}
    \|\sum_{r \in R} & \rho_r({}^{0}D^{\le e}) - \rho_r({}^{1}D^{\le e}) \|_1 = \|\sum_{r \in \hat R }\rho_r({}^{0}D^{\le e}) - \rho_r({}^{1}D^{\le e}) \|_1 \nonumber \\
    &\le \sum_{r \in \hat R } \Delta_{x} \rho_r\label{eq:actual_sensitivity}   
\end{align}

since ${}^{0}D^{\le e}$ and ${}^{1}D^{\le e}$ differ at most on $x = (d_0, e_0, F_0)$.    

Take $X^0 \sim X^1 \sim \lap(b)$ with $b = \sigma/\sqrt{2}$. We have:
\begin{align}
\label{eq:lapl}
   \frac{\Pr[V_{e,k}^0 = v_{e,k} | v_{<e,k}]}{\Pr[V_{e,k}^1 = v_{e,k} | v_{<e,k}]}  &= \frac{\Pr[\sum_{r \in R} \rho_r({}^{0}D^{\le e}) + X^0 = v_{e,k}]}{\Pr[\sum_{r \in R} \rho_r({}^{1}D^{\le e}) + X^1 = v_{e,k}]}
\end{align}

By property of the Laplace distribution, combining \Eq\ref{eq:actual_sensitivity} and \Eq\ref{eq:lapl} gives:

\begin{align}
  \left| \frac{\Pr[V_{e,k}^0 = v_{e,k} | v_{<e,k}]}{\Pr[V_{e,k}^1 = v_{e,k} | v_{<e,k}]}\right| \le \sum_{r \in \hat R }  \Delta_{x} \rho_r/b
\end{align}

By definition of $\computebudget$, we have $\epsilon_r = \Gamma_{x,r}/b$ where $\Delta_{x} \rho_r \le  \Gamma_{x,r}$.
Thus, we get $ \sum_{r \in \hat R } \Delta_{x} \rho_r/b \le  \sum_{r \in \hat R } \epsilon_r$.

Taking the sum over all queries, we get:

    \begin{align}
        \label{eq:gen2}
        |\ln \left(\frac{\Pr[V^0 = v]}{\Pr[V^1 = v]} \right)| & \le \sum_{e=1}^{e_{\max}} \sum_{k=1}^{k_{\max}} \sum_{r \in \hat R_{e,k} } \epsilon_r \\
        &\le \epsilon_x^G \label{eq:filter_conclusion}
    \end{align}

where \Eq\ref{eq:filter_conclusion} is by definition of a Pure DP filter.
\end{proof}

\begin{algorithm}[h!]
        \caption{Outer Privacy Game}
	\label{alg:outer_privacy_game}

	\begin{algorithmic}

        \Config
        \State Parametrized noise distribution $\cL$
        \State Device-epoch budget capacity $(\epsilon^G_x)_{x \in \cX}$
        \State Upper bound on number of epochs $e_{\max}$
        \State Upper bound on number of queries per epoch $k_{\max}$
        \State Public events $P \subset \cI \cup \cC$
        \EndConfig

        \Input
        \State Pair of records $x_0 = (d_0, e_0, F_0), x_1 = (d_1, e_1, F_1) \in \cX$ such that $e_0 = e_1$ and $F_0 \cap P = F_1 \cap P$
        \State Challenge bit $c$
        \State Adversary $\cB$
        \EndInput

        \Output
        \State View $W^c = (w_{1,0}^c, w_{1,1}^c, \dots, w_{1,k_{\max}}^c, w_{2,0}^c, w_{2,1}^c, \dots)$ of $\cB$
        \EndOutput
        \State\hrulefill

        \State Initialize \Alg\ref{alg:privacy_game} with same configuration, challenge bit $b=1$, opt-out device $x^c$ and adversary $\cA$ (whose behavior is defined next)
        \For{$e \in [e_{\max}]$}
        \State \graycomment{Generate data for the epoch $e$}
        \State Receive a database $G$ for epoch $e$ from $\cB$
        \State Ask $\cA$ to submit $G$
        \If{$e = e_0$ and $(d_0, e_0)  \not\in G$ and $(d_1, e_1)  \not\in G$} 
        \State \graycomment{At this point, $\cA$ also adds $x_c$ in his own game}
        \State $G^c \gets G + x_c$
        \Else
        \State $G^c \gets G$
        \EndIf

        \State \graycomment{Release public information}
        \State $S = \emptyset$
        \For{$(d,e,F) \in G^c$}
        \For{$f \in F \cap P $}
        \State Generate report nonce $r \overset{{\scriptscriptstyle\$}}{\leftarrow} U(\Z)$
        \State Save device corresponding to nonce $d_r \gets d$
        \State $S \gets S \cup \{(r,f)\}$
        \EndFor 
        \EndFor
        \State Send $w^c_{e,0} = S$ to $\cB$
        
        \State \graycomment{Answer queries after epoch $e$}
        \For{$k \in [k_{\max}]$}
        \State Receive query $Q_k$ from $\cB$ with corresponding nonces $R$, target epochs $(E_{r})_{r \in R}$, attribution functions $(A_r)_{r \in R}$ and noise std-dev $\sigma$.
        \State Ask $\cA$ to send $Q_k$ with devices $(d_r)_{r \in R}$, receive $(v_{x_c})^1_{e,k}$
        \State Send $w^c_{e,k} = (v_{x_c})^1_{e,k}$ to $\cB$
        \EndFor 
        \EndFor
        \end{algorithmic}
\end{algorithm}

\begin{theorem}[IDP of \Alg\ref{alg:outer_privacy_game} when replacing $x_0$ by $x_1$ for fixed public information]
    \label{thm:general_replace_guarantee}
    Fix a device-epoch budget capacity $(\epsilon^G_x)_{x \in \cX}$ for every possible record $x \in \cX$.
    Fix a set of public events $P \subset \cI \cup \cC$.
    
    For any pair of records $x_0 = (d_0, e_0, F_0), x_1 = (d_1, e_1, F_1) \in \cX$ such that $e_0 = e_1$ and $F_0 \cap P = F_1 \cap P$,
    for any adversary $\cB$, and $W^0,W^1$ defined by \Alg\ref{alg:outer_privacy_game},
    for all $w \in \operatorname{Supp}(W^1)$ we have:
            \begin{align}
            \label{eq:general_replace_guarantee}
             \left|\ln\left( \frac{\Pr[W^0 = v]}{\Pr[W^1 = v]} \right)\right| \le \epsilon^G_{x_0} + \epsilon^G_{x_1}     
            \end{align}
\end{theorem}
\begin{proof}
Fix an upper bound on the number of epochs and queries per epoch $e_{\max}, k_{\max}$.
Take a record pair $x_0,x_1 \in \cX$, an adversary $\cB$, $W^0,W^1$ defined by \Alg\ref{alg:outer_privacy_game} and $w \in \operatorname{Supp}(W^1)$.
We define $v := (w_{1,1}, \dots, w_{1,k_{\max}}, w_{2,1}, \dots, w_{e_{\max}, k_{\max}}$ the truncated version of the view $w$ without nonce information (steps with $k=0$).

We have:

\begin{align}
    \label{eq:prod_sep}
        \ln \left(\frac{\Pr[W^0 = w]}{\Pr[W^1 = w]} \right) & = \ln \left( \prod_{e=1}^{e_{\max}}\prod_{k=1}^{k_{\max}} \frac{\Pr[W_{e,k}^0 = w_{e,k} | w_{<e,k}]}{\Pr[W_{e,k}^1 = w_{e,k} | w_{<e,k}]} \right) \nonumber \\
        &+ \ln \left( \prod_{e=1}^{e_{\max}}\frac{\Pr[W_{e,0}^0 = w_{e,0} | v_{<e,0}]}{\Pr[W_{e,0}^1 = w_{e,0} | w_{<e,0}]} \right)
\end{align}

Take  $e \in [e_{\max}], k \in [k_{\max}]$, $c \in \{0,1\}$.
Take the database ${}^{c}D^{\le e}$ corresponding to $\cB$ conditioned on $w_{<e,k}$. 
$\cB$ receives two types of results:
\begin{itemize}
    \item If $k=0$, $W^c_{e,k}$ is about nonces and public events. 
    We denote by $Z$ the random variable that returns $\{ (U_f, f), f \in F\}$ with i.i.d. $U_f \sim \cU(\Z)$. 
    Since $F_0 \cap P = F_1 \cap P$, we have:

\begin{align}
\label{eq:uniform}
\Pr[W^0_{e,k} = w_{e,k} | w_{<e,k}] &= \Pr[Z = w_{e,k}] \nonumber \\ & = \Pr[W^1_{e,k} = w_{e,k} | w_{<e,k}]
\end{align}

    \item For $k > 0$, $W^c_{e,k}$ is the noisy answer to a query. 
    In \Alg\ref{alg:outer_privacy_game}, we instantiate $\cA$ as a valid adversary for \Alg\ref{alg:inner_privacy_game} with opt-out record $x_c$ and challenge bit $b=1$ (\ie $x_c$ is included in the database).
    We denote by $(V_{x_c})^{1}_{e,k}$ the view of this adversary $\cA$, and by definition of the truncated view $v$, we have:
    
    \begin{align}
    \label{eq:view_V}
        \Pr[W^c_{e,k} = w_{e,k} | w_{<e,k}] = \Pr[(V_{x_c})^{1}_{e,k} = v_{e,k} | v_{<e,k}] 
    \end{align}
\end{itemize} 

Thanks to \Eq\ref{eq:uniform} and \Eq\ref{eq:view_V}, \Eq \ref{eq:prod_sep} becomes:

\begin{align}
    \label{eq:Wloss}
    \ln &\left(\frac{\Pr[W^0 = w]}{\Pr[W^1 = w]} \right) \nonumber \\
    &= \ln \left(\frac{\Pr[V_{x_0}^1 = v]}{\Pr[V_{x_1}^1 = v]} \right) \nonumber \\
    &= \ln \left(\frac{\Pr[V_{x_0}^1 = v]}{\Pr[V_{x_1}^0 = v]} \right) + \ln \left(\frac{\Pr[V_{x_1}^0 = v]}{\Pr[V_{x_1}^1 = v]} \right)
\end{align}

We now show that $\Pr[V_{x_1}^0 = v] = \Pr[V_{x_0}^0 = v]$.
Take $e \in [e_{\max}], k \in [k_{\max}]$, and condition on a prefix $v_{<e,k}$.
Then, the only difference between $(V_{x_0})^0_{e,k}$  and $(V_{x_1})^0_{e,k}$ is the underlying database in \Alg\ref{alg:inner_privacy_game}, that we denote respectively $D$ and $D'$.
There exists a database $G$ such that ${}^0 D^{\le e} = G + \mathds{1}[e \le e_0] (d_0,e_0,\emptyset)$ and  ${}^0 D'^{\le e} = G + \mathds{1}[e \le e_1] (d_1,e_1,\emptyset)$.
Either way, for a report $\rho_r$ and a database $\D$, adding device-epoch records with empty events does not change the value of $\rho_r(D)$. Indeed, by definition $D_d^e$ already returns $\emptyset$ if $(d,e) \not \in D$.
Hence, $\sum_{r \in R} \rho_r({}^0 D^{\le e}) = \sum_{r \in R} \rho_r({}^0 D'^{\le e})  = \sum_{r \in R} \rho_r(G)$. 

Thus, 

\begin{align}
\label{eq:equivalent_empty_games}
\ln \left(\frac{\Pr[V_{x_0}^1 = v]}{\Pr[V_{x_1}^0 = v]} \right) =  \ln \left(\frac{\Pr[V_{x_0}^1 = v]}{\Pr[V_{x_0}^0 = v]} \right) 
\end{align}

Finally, by \Thm\ref{thm:inner_privacy}, \Eq\ref{eq:Wloss} becomes:
\begin{align}
    \left| \ln\left( \frac{\Pr[W^0 = v]}{\Pr[W^1 = v]} \right)\right| \le \epsilon^G_{x_0} + \epsilon^G_{x_1}  
\end{align}
\end{proof}

\begin{theorem}[Tighter \Thm\ref{thm:general_replace_guarantee} with constraint on queries]
    \label{thm:simplified_replace_with_public_dummy}
    Fix a set of public events $P \subset \cI \cup \cC$, and
    budget capacities $(\epsilon^G_x)_{x \in \cX}$.

    Take any $x = (d,e,F) \in \cX$, and define $x_P := (d,e,F \cap P)$.
    Suppose that all the attribution functions $A$ verify $\forall i, \forall F, \allowbreak\ A(F_1,\mydots,F_{i-1},F_i \cap P,\allowbreak F_i, \mydots, F_k) = A(F_1,\mydots,F_{i-1},\emptyset,F_i, \mydots, F_k)$.

    Then, for the record pair $(x, x_P)$,  for any adversary $\cB$, for $W^0,W^1$ defined by \Alg\ref{alg:outer_privacy_game} and
for all $w \in \operatorname{Supp}(W^1)$ we have:
            \begin{align}
             \left|\ln\left( \frac{\Pr[W^0 = v]}{\Pr[W^1 = v]} \right)\right| \le \epsilon^G_{x}     
            \end{align}

\end{theorem}
\begin{proof}
    First, we show that under such queries with  $F_A \cap P = \emptyset$, for any $x \in \cX$, \Alg\ref{alg:outer_privacy_game} produces the same output on $\epsilon^G_{x_P} = 0$ and $\epsilon^G_{x_P} > 0$.
    
    Take any $x = (d_0,e_0,F) \in \cX$, and define $x_P := (d_0,e_0,F \cap P)$.
    Take a report $\rho$ with an attribution function $A$ that is executed on $d_0$ and $E$ such that $e_0 \in E$.
    If  $\epsilon^G_{x_P} = 0$,  \Alg\ref{alg:outer_privacy_game} sets $F_{e_0} = \emptyset$ and returns $\rho = A((F_e)_{e \in E\setminus \{e_0\}} || \emptyset)$.
    If  $\epsilon^G_{x_P} > 0 $ and $\cF_{x_P}$ has enough budget, \Alg\ref{alg:outer_privacy_game} sets $F_{e_0} = F \cap P$ and returns $\rho = A((F_e)_{e \in E\setminus \{e_0\}} || F \cap P)$.
    Thanks to the constraint on $A$, we have $A((F_e)_{e \in E\setminus \{e_0\}} || \emptyset) = A((F_e)_{e \in E\setminus \{e_0\}} || F \cap P)$.

    Finally, we conclude with \Thm\ref{thm:general_replace_guarantee}.
\end{proof}

\subsection{Unlinkability Guarantees (\Thm\ref{thm:unlinkability})}

\begin{definition}[Unlinkability privacy game]
    \label{def:same_epoch_game}
    We define a variant of \Alg\ref{alg:outer_privacy_game} by applying the following modifications:
    \begin{itemize}
        \item We do not require $F_0 \cap P = F_1 \cap P$ anymore, and we define $x_2 := (d_0, e_0, F_0 \setminus F_1)$
        \item If $c=1$, after receiving $G$ from $\cB$, if $e = e_0$ and $x_2 \not \in G$, we perform $G \gets G + x_2$.
    \end{itemize}   
    In this variant, $\cB$ tries to distinguish between World 0 in which the database is $G + x_0 = G + (d_0, e_0, F_0)$, and World 1 in which the database is $G + x_1 + x_2 = G + (d_1, e_1, F_1) + (d_0, e_0, F_0 \setminus F_1)$.
    In World 0, all the events in $F_0$ are located on the same device, while in World 1 there are some events on device $d_0$ and some events on device $d_1$.
\end{definition}

\begin{theorem}[Unlinkability guarantees]
    \label{thm:general_unlinkability}
    Fix a set of public events $P \subset \cI \cup \cC$, and
    budget capacities $(\epsilon^G_x)_{x \in \cX}$.

    Take any $d_0, d_1 \in \cD$, $e \in \cE$, $F_0 \subset \cI \cup \cC$ and $F_1 \subset F_0$,
    and pose  $x_0 := (d_0, e, F_0), x_1 := (d_1, e, F_1), x_2 := (d_0, e, F_0 \setminus F_1) \in \cX$.
    Take any adversary $\cB$ for the game from \Def\ref{def:same_epoch_game} with record triple $(x_0, x_1, x_2)$, and note $U^0, U^1$ the views of $\cB$.

    Then, for all $u \in \operatorname{Supp}(U^1)$ we have:
    \begin{align}
     \left|\ln\left( \frac{\Pr[U^0 = u]}{\Pr[U^1 = u]} \right)\right| \le \epsilon^G_{x_0} + \epsilon^G_{x_1} +  \epsilon^G_{x_2}
    \end{align}

    This bounds the ability of $\cB$ to tell whether all the events $F_0$ (both public and private) belong to a single device or not.

\end{theorem}

\begin{proof}
    Take  $u \in \operatorname{Supp}(U^1)$.
    Similar to \Thm\ref{thm:general_replace_guarantee}, the nonce and public information follow the same distribution in $U^0$ and $U^1$, and the rest of the view corresponds to an execution of \Alg\ref{alg:inner_privacy_game} with challenge bit $b=1$.
    Hence we have:

    \begin{align}
    \label{eq:Uloss}
    \ln &\left(\frac{\Pr[U^0 = u]}{\Pr[U^1 = u]} \right) = \ln \left(\frac{\Pr[V_{x_0}^1 = v]}{\Pr[V_{x_1, x_2}^1 = v]} \right) \nonumber \\
    \end{align}

    where $u, V_{x_0}^1 , V_{x_1, x_2}^1$ are defined as follows:
    \begin{itemize}
        \item $v$ is the truncated version of $u$ obtained by removing the nonces and public information.
        \item $V_{x_0}^{b}$ is the view of the adversary $\cA$ defined in \Alg\ref{alg:outer_privacy_game} with $b \in \{0,1\}$, that if $b=1$ inserts the opt-out record $x_0$ in the database submitted by $\cB$.
        \item $V_{x_1, x_2}^b$ is the view of the adversary $\cA'$ defined in \Def\ref{def:same_epoch_game} with $b \in \{0,1\}$, that if $b=1$ inserts the opt-out record $x_1$ in the database submitted by $\cB$ extended with $x_2$.
        \item $V_{x_2}^b$ the view of the adversary $\cA"$ defined in \Alg\ref{alg:outer_privacy_game} with $b \in \{0,1\}$, that if $b=1$ inserts the opt-out record $x_2$ in the database submitted by $\cB$.

    \end{itemize}

    With the same reasoning as in \Thm\ref{thm:general_replace_guarantee} (\Eq\ref{eq:equivalent_empty_games}), we have $V_{x_0}^0 \sim V_{x_2}^0$.
    We also have $V_{x_1, x_2}^0 = V_{x_2}^1$. 
    Thus, \Eq\ref{eq:Uloss} becomes:

    \begin{align*}
    \ln &\left(\frac{\Pr[U^0 = u]}{\Pr[U^1 = u]} \right) = \ln \left(\frac{\Pr[V_{x_0}^1 = v]}{\Pr[V_{x_1, x_2}^1 = v]} \frac{\Pr[V_{x_1, x_2}^0 = v]}{\Pr[V_{x_2}^1 = v]} \frac{\Pr[V_{x_2}^0 = v]}{\Pr[V_{x_0}^0 = v]}  \right) 
    \end{align*}

    We conclude with \Thm\ref{thm:inner_privacy}.
\end{proof}

\begin{theorem}[Simplified Expression for \Thm\ref{thm:general_unlinkability}]
    \label{thm:simplified_unlinkability}
    Fix a set of public events $P \subset \cI \cup \cC$, and
    budget capacities $(\epsilon^G_x)_{x \in \cX}$.
    Take any $d_0, d_1 \in \cD$, $e \in \cE$, $F_1 \subset F_0 \subset P$ (\ie all the events we consider here are public events),
    and pose  $x_0 := (d_0, e, F_0), x_1 := (d_1, e, F_1), x_2 := (d_0, e, F_0 \setminus F_1) \in \cX$.
    Take any adversary $\cB$ for the game from \Def\ref{def:same_epoch_game} with record triple $(x_0, x_1, x_2)$,
    and note $U^0, U^1$ the views of $\cB$.
    Suppose that all the attribution functions $A$ submitted by $\cB$ have relevant events sets $I \cup C$ that verify $F_A \cap P = \emptyset$

    Then, for all $u \in \operatorname{Supp}(U^1)$ we have:
    \begin{align}
     \left|\ln\left( \frac{\Pr[U^0 = u]}{\Pr[U^1 = u]} \right)\right| = 0
    \end{align}
\end{theorem}

\begin{proof}
    First, we observe that $F_0 \cap F_A = F_1 \cap F_A = (F_0 \setminus F_1) \cap F_A = \emptyset$.
    Then, by applying the same reasoning as \Thm\ref{thm:simplified_replace_with_public_dummy}, we can suppose without loss of generality that $ \epsilon^G_{x_0} = \epsilon^G_{x_1} = \epsilon^G_{x_2} = 0$.
    We conclude with \Thm\ref{thm:general_unlinkability}.
\end{proof}

\subsection{Privacy Guarantees Under Colluding Queriers}
\label{sec:collusion}

We show that, as in DP, colluding parties can be analyzed using DP composition. 
This property is not immediate, because queriers in \sysname possess side information that they use to define queries with good IDP properties.
Informally, for a record $x$ on device $d$, the collusion of $n$ parties with budget $\epsilon^{G_1}_{d}, \dots, \epsilon^{G_n}_{d}$ is $2\epsilon^{G_1}_{d} + \dots + 2\epsilon^{G_n}_{d}$-DP for $x$ under the joint public information.
We can remove the factor 2 when queries never look at the public data from {\em any} colluding querier.

\begin{theorem}[Colluding Queriers]
    \label{thm:collusion}
    Fix $n > 1$ a number of colluding queriers (\ie adversaries from \Alg\ref{alg:outer_privacy_game}).
    For simplicity, we suppose that the data is not adaptively chosen, allowing us to see each querier as an interactive mechanism with view $\cM_i^{\leftrightarrow}(D)$ when executed on a database $D \in \D$.
    Fix a set of public events $P_i \subset \cI \cup \cC$ for each querier $i \in [n]$, and
    budget capacities $(\epsilon^{G_i}_x)_{x \in \cX}$.
    Define $P:= P_1 \cup \dots \cup P_n$.

    For any pair of records $x_0 = (d_0, e_0, F_0), x_1 = (d_1, e_1, F_1) \in \cX$ such that $e_0 = e_1$ and $F_0 \cap P = F_1 \cap P$,
    for any database $D \in \D$ with $(d_0,e_0) \not \in D, (d_1, e_1) \not \in D$,
    for any adversary $\cM$ that concurrently executes $\cM_1^{\leftrightarrow}, \dots, \cM_n^{\leftrightarrow}$ on the same data (potentially interleaving and adaptively choosing queries),
    for all $S \in \operatorname{Range}(\cM)$ we have:
            \begin{align}
             \Pr[\cM(D+x_0) \in S] \le \exp\left(\sum_{i=1}^n \epsilon^{G_i}_{x_0} + \epsilon^{G_i}_{x_1} \right) \Pr[\cM(D+x_1) \in S]
            \end{align}

    When the attribution functions used by any querier satisfy $\forall i, \forall F, \ A(F_1,\mydots,F_{i-1},F_i \cap P,F_i, \mydots, F_k) = A(F_1,\mydots,F_{i-1},\emptyset,F_i, \mydots, F_k)$, and when $x_1 = (d_0, e_0, F_0 \cap P)$, then we can remove the $\epsilon^{G_i}_{x_1}$ term.

\end{theorem}

In such a case of colluding queriers, the constraint that $\forall F, \ A(F \cap P) = A(\emptyset)$ is more restrictive than merely asking $\forall F, A^i(F \cap P_i) = \emptyset$ for a single querier as in \Thm\ref{thm:simplified_replace_with_public_dummy}. 
For instance, the queries we describe in \S\ref{sec:data-query-model-instantiations} will not verify this constraint if an advertiser and a publisher collude. However, the guarantee under general queries of $2\sum_{i=1}^n\epsilon^{G_i}_d$-DP still applies.

\begin{proof}
    The key observation is that \Thm\ref{thm:general_replace_guarantee} shows that \Alg\ref{alg:outer_privacy_game} is in particular DP under a more restrictive \changeone neighborhood relation over the union of the public information across queriers.
    We can then compose $n$ mechanisms under this restrictive neighborhood relation. 
    
    More formally, fix $Q \subset \cI \cup \cC$ and $x = (d,e,F), x'=(d', e', F') \in \cX$.
    We define the following neighborhood relation over databases.
    For $D,D' \in \D$, we say $D \overset{{\scriptscriptstyle Q}}{\underset{\scriptscriptstyle x,x'}{\sim}} D$ if $e = e'$, $F \cap Q = F' \cap Q$, and there exists $D_0 \in \D$ such that $D = D_0 + x$ and $D' = D_0 + x'$ or vice versa. 
    Consider $x_0 = (d_0, e_0, F_0), x_1 = (d_1, e_1, F_1) \in \cX$ such that $e_0 = e_1$.
    For all $i \in [n]$, we have $F_0 \cap P = F_1 \cap P \implies F_0 \cap P_i = F_1 \cap P_i$, %
    and thus:
    \begin{align}
    \label{eq:neighborhood_implication}
    \forall D,D' \in \D,  D \overset{{\scriptscriptstyle P}}{\underset{\scriptscriptstyle x_0,x_1}{\sim}} D  \implies D \overset{{\scriptscriptstyle P_i}}{\underset{\scriptscriptstyle x_0,x_1}{\sim}} D
    \end{align}
    
    \Thm\ref{thm:general_replace_guarantee} shows the interactive mechanism $\cM_i^{\leftrightarrow}$ is $\epsilon^{G_i}_{x_0} + \epsilon^{G_i}_{x_1}$-DP under the $\overset{{\scriptscriptstyle P_i}}{\underset{\scriptscriptstyle x_0,x_1}{\sim}}$ relation.
    Thanks to \Eq\ref{eq:neighborhood_implication},  $\cM_i^{\leftrightarrow}$ is also $\epsilon^{G_i}_{x_0} + \epsilon^{G_i}_{x_1}$-DP under the $\overset{{\scriptscriptstyle P}}{\underset{\scriptscriptstyle x_0,x_1}{\sim}}$ relation.
    Note that this conclusion would not be true if we had proved \Thm\ref{thm:general_replace_guarantee} under the replace-with-default definition $D \sim_x^Q D'$ introduced in \S \ref{sec:data-model}.

    Next, the adversary that concurrently executes the $n$ queriers is operating a concurrent composition of interactive mechanisms $\cM_1^{\leftrightarrow}, \dots, \cM_n^{\leftrightarrow}$.
    Thanks to the concurrent composition theorem \cite{VZ23}, the resulting mechanism $\cM$ is $\sum_{i=1}^n \epsilon^{G_i}_{x_0} + \epsilon^{G_i}_{x_1}$-DP under the $\overset{{\scriptscriptstyle P}}{\underset{\scriptscriptstyle x_0,x_1}{\sim}}$ relation.

\end{proof}

%% file: appendix/04-idp-optimizations-proofs.tex
\section{Proofs for IDP Optimizations (\S\ref{sec:idp-optimizations})}
\label{sec:appendix:idp-optimizations-proofs}

\begin{theorem}[Global sensitivity of reports]
	\label{thm:apx:global_sensitivity_of_reports}
	Fix a report identifier $r$, a device $d_r$, a set of epochs $E_r$, an attribution function $A$ and the corresponding report $\rho: D \mapsto A(D^{E_r}_{d_r})$.
	We have:
	\begin{align*}
		\Delta (\rho) = \underset{i \in [k], F_1, \dots, F_k \in \cP(\cI \cup \cC)}{\max \|A(F_1, \mydots, F_k) } - A(F_1, \mydots, F_{i-1}, \emptyset, F_{i+1}, \mydots, F_k) \|_1
	\end{align*}

        If $A$ has $m$-dimensional output and verifies $\forall \mathbf{F} \in \cP(\cI \cup \cC)^k, \forall i \in [m], A(\mathbf{F})_i \in [0,A^{\max}]$, then we have $\Delta(\rho) \le m A^{\max}$. 
\end{theorem}
\begin{proof}
    Take such a report $\rho$. 
    We enumerate the requested epochs from $1$ to $k=|E_r|$: $E_r =\{e_1, \dots, e_k\}$.
    
    First, by definition of the global sensitivity, we have:

    \begin{align}
        \Delta (\rho) &= \max_{D,D' \in \D: \exists x \in \cX, D' = D + x } \|\rho(D) - \rho(D')\|_1 \\
        &= \max_{D,D' \in \D: \exists x \in \cX, D' = D + x } \|A(D^{E_r}_{d_r}) - A((D')^{E_r}_{d_r})\|_1\\
        &= \max_{D,D' \in \D: \exists x = (d_r, e, F) \in \cX : e \in E_r, D' = D + x } \|A(D^{E_r}_{d_r}) - A((D')^{E_r}_{d_r})\|_1
    \end{align}

    since for $x = (d,e,F)$ with $d \neq d_r$ or $e_r \not \in E_r$ we have $A(D^{E_r}_{d_r}) = A((D')^{E_r}_{d_r})$.

    Next, we show that the two following sets are equal:
    \begin{itemize}
        \item $\{ (D^{E_r}_{d_r}), (D')^{E_r}_{d_r}) | D,D' \in \D: \exists x = (d_r, e, F) \in \cX : e \in E_r, D' = D + x\}$
        \item $\{ ((F_1, \mydots, F_{i-1}, \emptyset, F_{i+1}, \mydots, F_k), (F_1, \mydots, F_k) ) | i \in [k], F_1, \newline \dots, F_k \in \cP(\cI \cup \cC) \}$
    \end{itemize}

    On one hand, take $D,D' \in \D$ such that there exists $x = (d_r, e, F) \in \cX$ verifying $e_r \in E_r$ and $D' = D + x$.
    We pose $F_j := (D')_{d_r}^{e_j}$ for $e_j \in E_r$.
    If $x$ has epoch $e = e_i \in E_r$ for some $i$, then we have $F_i = F$.
    Hence, since $D$ must not contain $(d_r, e)$, we have:
    $D^{E_r}_{d_r} = (F_1, \mydots, F_{i-1}, \emptyset, F_{i+1}, \mydots, F_k)$ and $(D')^{E_r}_{d_r} = (F_1, \mydots, F_k)$.

    Reciprocally, take $F_1, \mydots, F_k \in \cP(\cI \cup \cC)$ and $i \in [k]$.
    We define $D' := \{(d_r, e_1, F_1), \dots, (d_r, e_k, F_k)\}$ and $D' := D \setminus (d_r, e_i, F_i)$.
    We have $D,D' \in \D$ and there is $x = (d_r, e_i, F_i) \in \cX$ such that $D' = D + x$.

    Thus both sets are equal, and the maximum becomes:

    \begin{align}
        		\Delta (\rho) = \underset{i \in [k], F_1, \dots, F_k \in \cP(\cI \cup \cC)}{\max \|A(F_1, \mydots, F_k) } - A(F_1, \mydots, F_{i-1}, \emptyset, F_{i+1}, \mydots, F_k) \|_1
    \end{align}

    Finally, suppose that $A$ has output in $\R^m$. Take $\mathbf{F}, \mathbf{F'}$.
    We have $\|A(\mathbf{F}) - A(\mathbf{F'})\|_1 = \sum_{i=1}^m |A(\mathbf{F})_i - A(\mathbf{F'})_i|$.
    For $i \in [m]$ we have $A(\mathbf{F})_i  \in [0, A^{\max}]$ so $A(\mathbf{F})_i - A(\mathbf{F'})_i \in [-A^{\max}, A^{\max}]$.
    Hence $\|A(\mathbf{F}) - A(\mathbf{F'})\|_1 \le m A^{\max}$.

    This upper bound on $\Delta(\rho)$ can be refined if $A$ has certain properties, such as being a histogram query.    
\end{proof}

\begin{theorem}[Global sensitivity of queries]
	\label{thm:apx:global_sensitivity_of_queries}

	Fix a query $Q$ with corresponding report identifiers $R$ and reports, devices and epoch windows $(\rho_r, d_r, E_r)_{r\in R}$.
	\begin{align}
		\Delta(Q) \le \max_{d,e} \sum_{r \in R: d = d_r, e \in E_r} \Delta(\rho_r)
	\end{align}

	In particular, if each device-epoch participates in at most one report, then $\Delta(Q) = \max_{r \in R} \Delta(\rho_r)$.

\end{theorem}
\begin{proof}
    Take such a query $Q$.
    We observe that 
    
    \begin{align}
        \Delta(Q)  &= \max_{D,D' \in \D: \exists x \in \cX, D' = D + x } \|Q(D) - Q(D')\|_1 \\ 
        &= \max_{x \in \cX} \max_{D,D' \in \D: D' = D + x } \|Q(D) - Q(D')\|_1 
    \end{align} 

    Take $x = (d,e,F) \in \cX$. 
    For $r \in R$ such that $d \neq d_r$ or $e \not \in E_r$, we have $\rho_r(D) = \rho_r(D')$. Thus:

    \begin{align}
        \|Q(D) - Q(D')\|_1 & = \| \sum_{r \in R} \rho_r(D) - \rho_r(D') \|_1 \\
        & = \| \sum_{r \in R: d = d_r, e \in E_r} \rho_r(D) - \rho_r(D') \|_1 
    \end{align}

    Using the triangle inequality and the definition of $\Delta(\rho)$ we get:
    \begin{align}
        \|Q(D) - Q(D')\|_1 &\le \sum_{r \in R: d = d_r, e \in E_r} \|\rho_r(D) - \rho_r(D') \|_1 \\
        &\le \sum_{r \in R: d = d_r, e \in E_r} \Delta(\rho_r)
    \end{align}

    This bound is independent on $D,D'$ so:
    
    \begin{align}
        \max_{D,D' \in \D: D' = D + x }  \|Q(D) - Q(D')\|_1  \le \sum_{r \in R: d = d_r, e \in E_r} \Delta(\rho_r) 
    \end{align}

    Finally, this does not involve $F$ so we can replace the max over $x = (d,e,F)$ by a max over $(d,e)$:

    \begin{align}
        \max_{x \in \cX} \max_{D,D' \in \D: D' = D + x } \|Q(D) - Q(D')\|_1 \le \max_{d,e} \sum_{r \in R: d = d_r, e \in E_r} \Delta(\rho_r)
    \end{align}

    If each device-epoch participates in at most one report, then this becomes $\Delta(Q) \le \max_r \Delta(\rho_r)$.
    For each $r$ there exists a pair $D,D'$ such that $\|\rho_r(D) - \rho_r(D')\|_1 = \Delta(\rho_r)$.
    Taking the max across reports shows that the upper bound on $\Delta(Q)$ is tight in this case.
    
\end{proof}

\begin{theorem}[Individual sensitivity of reports]
	\label{thm:apx:individual_sensitivity_of_reports}
	Fix a report identifier $r$, a device $d_r$, a set of epochs $E_r$, an attribution function $A$ with relevant events $F_A$, and the corresponding report $\rho: D \mapsto A(D^{E_r}_{d_r})$.
	Fix a device-epoch record $x = (d,e,F) \in \cX$.

	If the report requests a single epoch $E_r = \{e_r\}$, we have:
	\begin{align}
		\Delta_x(\rho) =
		\begin{cases}
			\|A(F) - A(\emptyset)\|_1 & \text{if } d = d_r \text{ and } e = e_r \\
			0                 & \text{otherwise}
		\end{cases}
	\end{align}

	Otherwise, we have:

	\begin{align}
		\Delta_x(\rho) \le
		\begin{cases}
			\Delta(\rho) & \text{if } d = d_r \text{ and } e \in E_r \text{ and } F \cap F_A \neq \emptyset \\
			0          & \text{otherwise}
		\end{cases}
	\end{align}
\end{theorem}
\begin{proof}
    Fix such a report $\rho$ and $x \in (d,e,F) \in \cX$.
    Consider any $D,D' \in \D$ such that $D' = D + x$.
    We have $\rho(D) = A(D_{d_r}^{e_r})$ and $\rho(D') = A((D')_{d_r}^{e_r})$

    \begin{itemize}
        \item
    First, suppose that the report requests a single epoch $e_r$. 

    \begin{itemize}
        \item If $d = d_r$ and $e = e_r$, then since $D + x \in \D$ we must have $(d_r,e_r) \not \in D$, and thus $D_{d_r}^{e_r} = \emptyset$. On the other hand, we have $(D')_{d_r}^{e_r} = F$. Thus, $\|\rho(D) - \rho(D')\|_1 = \|A(F) - A(\emptyset)\|_1$
        \item If $d \neq d_r$ or $e \neq e_r$, then $(D')_{d_r}^{e_r} = D_{d_r}^{e_r}$. Hence $\|\rho(D) - \rho(D')\|_1 = 0$.
    \end{itemize}
    These equalities are independent on $D,D'$, so taking the max gives $\Delta_x(\rho) = \|A(F) - A(\emptyset)\|_1$ if $d = d_r$ and $e = e_r$, and $0$ otherwise.

    \item Second, suppose that the report requests an arbitrary range of epochs $E_r$.
    \begin{itemize}
        \item If $d \neq d_r$ or $e \neq E_r$, then $(D')_{d_r}^{E_r} = D_{d_r}^{E_r}$. Hence $\|\rho(D) - \rho(D')\|_1 = 0$.
        \item If $d = d_r \text{ and } e = e_i \in E_r$ and $F \cap F_A = \emptyset$, we have $(D')_{d_r}^{E_r} = (D_{d_r}^{e_1}, \mydots, D_{d_r}^{e_{i-1}}, F, D_{d_r}^{e_{i+1}}, \mydots, D_{d_r}^{e_k})$. By definition of $I_A \cup C_A$, we have $A((D')_{d_r}^{E_r} ) = A(D_{d_r}^{e_1} \cap F_A, \mydots, D_{d_r}^{e_{i-1}}\cap F_A, F\cap F_A, D_{d_r}^{e_{i+1}}\cap F_A, \mydots, D_{d_r}^{e_k}\cap F_A) $.
        
        We also have $D_{d_r}^{E_r} = (D_{d_r}^{e_1}, \mydots, D_{d_r}^{e_{i-1}}, \emptyset, D_{d_r}^{e_{i+1}}, \mydots, D_{d_r}^{e_k})$.
        Since $F\cap F_A = \emptyset = \emptyset \cap F_A$, we get $A((D')_{d_r}^{E_r} )  = A(D_{d_r}^{E_r} ) $ \ie $\|\rho(D) - \rho(D')\|_1 = 0$.
        
        \item Otherwise, we must have $ d = d_r \text{ and } e \in E_r \text{ and } F \cap F_A \neq \emptyset$.
        In that case, $\|\rho(D) - \rho(D')\|_1 = \| A(D_{d_r}^{e_1}, \mydots, \\
        D_{d_r}^{e_{i-1}}, F, D_{d_r}^{e_{i+1}}, \mydots, D_{d_r}^{e_k}) - A((D_{d_r}^{e_1}, \mydots, D_{d_r}^{e_{i-1}}, \emptyset, D_{d_r}^{e_{i+1}}, \newline \mydots, D_{d_r}^{e_k}))\|_1$.
    \end{itemize}

    The first two identities are independent on $D,D'$, so taking the max gives $\Delta_x(\rho) = 0$.
    Unfortunately, the third identity depends on $D,D'$. Taking the max gives:

    \begin{align*}
         \Delta_x(\rho)  &=  \underset{F_1, \mydots, F_{i-1}, \emptyset, F_{i+1}, \mydots, F_k \in \cP(\cI \cup \cC)}{\max  \|A(F_1, \mydots, F_{i-1}, F,} F_{i+1}, \mydots, F_k)\\
        & \hspace{3mm} - A(F_1, \mydots, F_{i-1}, \emptyset, F_{i+1}, \mydots, F_k) \|_1 \\
        & \le \underset{i \in [k], F_1, \dots, F_k \in \cP(\cI \cup \cC)}{\max  \|A(F_1, \mydots, F_{i-1}}, F, F_{i+1}, \mydots, F_k)\\
        & \hspace{3mm} - A(F_1, \mydots, F_{i-1}, \emptyset, F_{i+1}, \mydots, F_k) \|_1 \\
        &= \Delta(\rho)
    \end{align*}

    thanks to \Thm\ref{thm:apx:global_sensitivity_of_reports}.
    Although we can technically keep the first equality to get a tighter expression for $\Delta_x(\rho)$, for common attribution functions $\Delta(\rho)$ is just as tight (\eg if the attribution cap is attained because of another possible epoch $F_j, j \neq i$).

    \end{itemize}

\end{proof}

\begin{theorem}[Individual sensitivity of queries]
	\label{thm:apx:individual_sensitivity_of_queries}

	Fix a query $Q$ with corresponding report identifiers $R$ and reports $(\rho_r)_{r\in R}$.
	Fix a device-epoch record $x = (d,e,F) \in \cX$.
	We have:

	\begin{align}
		\Delta_x (Q) \le \sum_{r \in R} \Delta_x(\rho_r)
	\end{align}

	In particular, if $x$ participates in at most one report $\rho_r$, then $\Delta_x(Q) = \Delta_x(\rho_r)$.
\end{theorem}

\begin{proof}
    The inequality is immediate by triangle inequality and definition of individual sensitivity.
    When $x$ participates in at most one report $\rho_r$, we get $\Delta_x(\rho_{\hat r}) = 0$ for $\hat r \neq r$, and thus $\Delta_x(Q) \le \Delta_x(\rho_r)$.
    The inequality is tight in that case.
\end{proof}

%% file: appendix/05-idp-bias-detection.tex
\section{IDP-Induced Bias Detection}
\label{sec:appendix:bias-detection}

Since individual privacy budgets depend on the data, they must be kept private. 
That is why \sysname silently replaces out-of-budget device-epoch data by $\emptyset$ instead of raising an exception like \IPA.
This missing data induces a bias in the query answers and increases the overall error.

\heading{IDP-induced bias.}
Consider a query $Q$ with report identifiers $R$, target epochs $(E_{r})_{r \in R}$, attribution functions $(A_r)_{r \in R}$ and noise parameter $\sigma$.
For a database $D$, the true result is $Q(D)  = \sum_{r \in R} A_r(D^{E_r}_{d_r})$.
When a device-epoch $(d_r, e)$ is out of budget, \sysname drops it, \ie \Alg\ref{alg:e2e_functional_view} uses $F_e = \emptyset$ instead of $F_e = D^{e}_{d_r}$.
We pose:

\begin{align}
\tilde Q(D) := \sum_{r \in R} A_r((F_e)_{e \in E_r})
\label{eq:q_tilde}
\end{align}

We denote by $\cM(D)$ the value returned by $\operatorname{AnswerQuery}$: $\cM(D) := \tilde Q(D) + X$ where $X \sim \cL(\sigma)$ has mean zero and variance $\sigma^2$.
Hence, \Alg\ref{alg:e2e_functional_view} returns an estimate for $Q(D)$ with the following bias:
\begin{align}
    \label{eq:bias_base}
    \E[ \cM(D) - Q(D)] = \tilde Q(D) - Q(D)
\end{align}

\heading{Detecting bias with global sensitivity.}
When no device-epoch is out of budget, \Alg\ref{alg:e2e_functional_view} returns an unbiased estimate.
We can guarantee that no device-epoch is out of budget by keeping track of a budget consumption bound as follows.
Assume we know 
(1) a lower bound $\epsilon^G$ on the individual budget capacity: $\forall x \in D, \epsilon^G_x \ge \epsilon^G$, and 
(2) an upper bound on the individual budget for each report $r$ in each query $k$: $\epsilon_x^{k,r} \le \epsilon^{k,r}$.
Then, for all $x \in D$, $\sum_{k,r} \epsilon^{k,r} \le \epsilon^G \implies \sum_{k,r} \epsilon_x^{k,r} \le \epsilon^G_x$.

In practice, the individual budget can be bounded by using the fact that the individual sensitivity is upper bounded by the (data-independent) global sensitivity.
Hence, a querier can run its own off-device budgeting scheme to detect the earliest potentially biased query.
This approach does not consume any budget since it only relies on public query information. 
However, once $\sum_{k,r} \epsilon^{k,r} > \epsilon^G$ this approach doesn't guarantee that queries are biased (or unbiased). 

\heading{Estimating bias with DP counting.}
To get a more granular estimate of the bias, we can run a special query counting the  number of reports that contain an out-of-budget epoch, as follows.
Given a query $Q$ with output in $\R^m$, we atomically execute $(Q_0, Q)$ as a single query with output in $\R^{m+1}$, where 
$Q_0(D) := \sum_{r \in R} \kappa \cdot \mathds{1}[\exists e \in E_r: D^e_{d_r} = \emptyset]$.
Prepending a side query to $Q$ gives a high probability bound on the bias.

\heading{Results.} \Thm\ref{thm:bias_whole_report} formally states the general high probability bound on the bias described above. \Thm\ref{thm:bias_whole_report_last_touch} specializes \Thm\ref{thm:bias_whole_report} to last-touch attribution.
These side queries increase the privacy budget, as stated in \Thm\ref{thm:bias_whole_report_sensitivity}.
Finally, \Thm\ref{thm:histogram_attribution_function} shows that expressions in \Thm\ref{thm:bias_whole_report} and \Thm\ref{thm:bias_whole_report_last_touch} can use tighter bounds from  for certain common attribution functions.    

\begin{theorem}
    \label{thm:bias_whole_report}
    Take a query $Q$ with report identifiers $R$, parameters $(d_r,E_r,A_r, \rho_r)_{r \in R}$, and output in $\R^m$.
    Fix $\kappa > 0$, a parameter to control the precision of the bound.
    For $r \in R$, we define $\hat A_r:\cP(\cI \cup \cC) \to \R^{m+1}$ by:
    $\hat A_r( F_1, \dots, F_k)_0 = \kappa \cdot \sum_{i=1}^k \mathds{1}[F_i = \emptyset]$ and $\forall i \in [m+1], \hat A_r( F_1, \dots, F_k)_i =  A_r( F_1, \dots, F_k)_i$.
    We pose $Q_0(D) := \sum_{r \in R} \kappa \cdot \mathds{1}[\exists e \in E_r: D^e_{d_r} = \emptyset]$, and denote by $(\cM_0(D), \cM(D))$ the output of \Alg\ref{alg:e2e_functional_view} on $(Q_0, Q)$, using Laplace noise with standard deviation $\sigma$.

    For a report $\rho$ with attribution function $A$ over $k$ epochs, we also define:

    \begin{align}
        \Delta^{\max}(\rho_r) := \max_{\mathbf{F},\mathbf{F}' \in \cP(\cI \cup \cC)^k: \forall i \in [k], \mathbf{F}_i' = \mathbf{F}_i \text{ or } \mathbf{F}_i' = \emptyset} \|A(\mathbf{F}) - A(\mathbf{F}')\|_1
    \end{align}

    That is, $\Delta^{\max}(\rho_r)$ is the maximum L1 change that happens in a report when we remove {\em any} number of epochs from a device.
    By comparison, the global sensitivity $\Delta(\rho_r)$ is the maximum change that happens when we remove {\em a single} device-epoch from the database.
    For certain attribution functions, such as last touch attribution, $\Delta^{\max}(\rho_r) = \Delta(\rho_r)$, as detailed next in \Thm~\ref{thm:histogram_attribution_function}.

    Then, for any $\beta \in (0,1)$, with probability $1-\beta$ we have: 

    \begin{align*}
        \left\Vert\E[\cM(D) - Q(D)]\right\Vert_1 \le  \frac{\cM_0(D) + \sigma \ln(1/\beta)/\sqrt{2}}{\kappa} \max_{r\in R} \Delta^{\max}(\rho_r)
    \end{align*}    
    
\end{theorem}

\begin{proof}
    First, we remark that $\cM_0(D)$ is an  {\em unbiased} estimate of  $\tilde Q_0(D)$, by definition of $\tilde Q_0$ which is the output of $Q_0$ after dropping out-of-budget epochs from $D$. 
    $\tilde Q_0(D)$ can then be used to get an upper bound on the number of reports containing at least one out-of-budget epoch.
    Indeed, when $d_r,e$ runs out of budget, $\hat A_r$ receives $F_{r,e} = \emptyset$ in \Alg\ref{alg:e2e_functional_view}. 
    Hence:
    
    \begin{align}
        \tilde Q_0(D)/\kappa &= |\{ r \in R: \exists e \in E_r, F_{r,e} = \emptyset \}|\\
        &= |\tilde R| \label{eq:q_tilde_bound} 
    \end{align}

where $\tilde R := \{ r \in R: \exists e \in E_r, F_{r,e} = \emptyset \}|$.

    Second, we can use $\tilde Q_0(D)$ to bound the bias as follows:
    \begin{align}
        \label{eq:bias2}
        &\left\Vert \E[ \cM(D)  - Q(D)]\right\Vert_1 = \|\tilde Q(D) - Q(D) \|_1  \\
        &= \|\sum_{r \in R} A(F_{r,e_1}, \dots, F_{r,e_k}) - A(D_{d_r}^{e_1}, \dots, D_{d_r}^{e_k} \|_1 \\
        &\le \sum_{r \in R} \|A(F_{r,e_1}, \dots, F_{r,e_k}) - A(D_{d_r}^{e_1}, \dots, D_{d_r}^{e_k} \|_1 \\
        &= \sum_{r \in R: A(F_{r,e_1}, \dots, F_{r,e_k}) \neq A(D_{d_r}^{e_1}, \dots, D_{d_r}^{e_k})}\|A(F_{r,e_1}, \dots, F_{r,e_k}) \\
        & \hspace{4.5cm} - A(D_{d_r}^{e_1}, \dots, D_{d_r}^{e_k} \|_1 \nonumber
    \end{align}

    The set of altered reports $\{ r \in R: A(F_{r,e_1}, \dots, F_{r,e_k}) \neq A(D_{d_r}^{e_1}, \dots, D_{d_r}^{e_k})\}$ is not directly accessible through our counting query, but it is a subset of the set of reports containing empty epochs.
    These two sets are not necessarily equal, because certain epochs could be empty in the original database (unless the application programmatically enforces $D_{d_r}^e \neq \emptyset$ by adding a special heartbeat event $f_0 \in F$ in every active device-epoch), and some out-of-budget epochs can leave the final report value unchanged. In other words:
    
    \begin{align}\{ r \in R: A(F_{r,e_1}, \dots, F_{r,e_k}) \neq A(D_{d_r}^{e_1}, \dots, D_{d_r}^{e_k}\} \subset \tilde R
    \label{eq:altered_report_subset}
    \end{align}

Hence, we have:
    \begin{align}
    \left\Vert \E[ \cM(D)  - Q(D)]\right\Vert_1 &\le \sum_{r \in \tilde R }\|A(F_{r,e_1}, \dots, F_{r,e_k}) - A(D_{d_r}^{e_1}, \dots, D_{d_r}^{e_k} \|_1\\
        &\le |\tilde R| \max_{r\in R}\Delta^{\max}(\rho_r) \\
        &\le (\tilde Q_0(D)/\kappa) \max_{r\in R} \Delta^{\max}(\rho_r) \label{eq:bias_bound_by_q0}
    \end{align}
    thanks to \Eq\ref{eq:q_tilde_bound}.

    Finally, we can use a tail bound to get a high probability bound on the expected bias.
    The knob $\kappa$ controls the precision of the out-of-budget count: higher $\kappa$ gives a more precise estimate but consumes more budget.
    More precisely, for Laplace noise with standard deviation $\sigma$, for an absolute error $\tau = \frac{\sigma \ln(1/\beta)}{\kappa\sqrt{2}}$ in the number of potentially biased reports and a failure probability target $\beta \in (0,1)$,
    we have:
    
    \begin{align}
        \label{eq:tail_bound_for_bias_count}
        \Pr[ | \cM_0(D)/\kappa - \tilde Q_0(D) /\kappa | > \tau ] = \beta
    \end{align}

    We conclude by injecting \Eq\ref{eq:tail_bound_for_bias_count} into \Eq\ref{eq:bias_bound_by_q0}.
    
\end{proof}

\begin{theorem}
    \label{thm:bias_whole_report_last_touch}
    Consider the setting defined in \Thm\ref{thm:bias_whole_report}.
    Additionally, suppose that $A$ performs last touch attribution, where epochs identifiers $\cE \subseteq \N$ are ordered chronologically.
    We replace $Q_0$ by the following counting query:
    $Q_0(D) := \sum_{r \in R} \kappa \cdot \mathds{1}[\exists i \in E_r: D^{i}_{d_r} = \emptyset \wedge \forall j \in E_r : j > i, D^j_{d_r} \cap F_A = \emptyset]$.

    Then, we also have: 

    \begin{align*}
        \left\Vert\E[\cM(D) - Q(D)]\right\Vert_1 \le  \frac{\cM_0(D) + \sigma \ln(1/\beta)/\sqrt{2}}{\kappa} \max_{r\in R} \Delta^{\max}(\rho_r)
    \end{align*}    
    
\end{theorem}
\begin{proof}

    The only difference with \Thm\ref{thm:bias_whole_report} is that \Eq\ref{eq:q_tilde_bound} becomes:
    \begin{align}
        \tilde Q_0(D)/\kappa &= |\tilde R| \label{eq:new_q_tilde_bound} 
    \end{align}

where $\tilde R := \{ r \in R: \exists i \in E_r, F_{r,i} = \emptyset \wedge \forall j \in E_r : j > i, D^j_{d_r} \cap F_A = \emptyset\}|$.
    Importantly, this new definition of $\tilde R$ still verifies the same identity as \Eq\ref{eq:altered_report_subset}:

    \begin{align}\{ r \in R: A(F_{r,e_1}, \dots, F_{r,e_k}) \neq A(D_{d_r}^{e_1}, \dots, D_{d_r}^{e_k})\} \subset \tilde R
    \end{align}

    Indeed, let's show that $A(F_{r,e_1}, \dots, F_{r,e_k}) \neq A(D_{d_r}^{e_1}, \dots, D_{d_r}^{e_k}) \newline \implies r \in \tilde R$.
    Consider a report $r \in R \setminus \tilde R$.
    By definition of $\tilde R$ we have for all $i \in E_r$,
    
    \begin{align}
        \label{eq:report_marked_as_unbiased}
        F_{r,i} \neq \emptyset \vee \exists j > i : F_j \cap F_A \neq \emptyset
    \end{align}

        We now use the assumption that $A$ performs last-touch attribution to show $A(F_{r,e_1}, \dots, F_{r,e_k}) = A(D_{d_r}^{e_1}, \dots, D_{d_r}^{e_k})$.
    \begin{itemize}
        \item If $D_{d_r}^{e_1}, \dots, D_{d_r}^{e_k}$ contain no attributable impressions, then $A(F_{r,e_1}, \dots, F_{r,e_k}) = A(D_{d_r}^{e_1}, \dots, D_{d_r}^{e_k})$.
        \item Otherwise, denote by $i$ the epoch containing the last-touch, \ie the most recent relevant event $f \in F_A$, that should get full attribution. 
        By definition of $i$, $\forall j > i, D_{d_r}^{j} \cap F_A = \emptyset$. But since $r \in R \setminus \tilde R$, \Eq\ref{eq:report_marked_as_unbiased} implies $F_{r,i} \neq \emptyset$.
        Thus $F_{r,i} = D_{d_r}^{i}$, and since more recent epochs $j>i$ do not contain relevant events, the full attribution value is allocated to the same event in both cases: $A(F_{r,e_1}, \dots, F_{r,e_k}) = A(D_{d_r}^{e_1}, \dots, D_{d_r}^{e_k})$.
    \end{itemize}
    
    The rest of the proof is identical.
\end{proof}

\begin{theorem}[Sensitivity of augmented queries]
    \label{thm:bias_whole_report_sensitivity}
    Consider a query  $(d_r,E_r,A_r, \rho_r)_{r \in R}$ augmented by a side query such that each report $\hat \rho_r : D \mapsto (\rho_r^0(D), \rho_r(D)) \in \R^{m+1}$ verifies  $\rho_r^0(D) \in [0, \kappa]$ for some fixed $\kappa >0$.

    Take $x = (d,e,F) \in \cX$. We have:
        \begin{align}
        \Delta_x(\hat \rho_r) \le \kappa \cdot \mathds{1}[d = d_r, e \in E_r \text{ and } F \neq \emptyset] + \Delta_x(\rho_r)
    \end{align}

\end{theorem}
\begin{proof}
        First, we have:

    \begin{align}
        \Delta_x(\hat \rho_r) \le \Delta_x(\rho_r^0) + \Delta_x(\rho_r)
    \end{align}

    because for all $D,D'$ such that $D' = D + x$ we have $\|\hat \rho_r(D') - \hat \rho_r(D)\|_1 \le \| \rho_r^0(D') -  \rho_r^0(D)\|_1 + \|\rho_r(D') - \rho_r(D)\|_1 \le   \Delta_x(\rho_r^0) + \Delta_x(\rho_r)$.

    Second, we have:
    \begin{align}
    \label{eq:sensitivity_count}
	\Delta_x(\rho_r^0) \le	\begin{cases}
			 \kappa &  \text{if } d = d_r, e \in E_r \text{ and } F \neq \emptyset \\
			0                 & \text{otherwise}
		\end{cases}
    \end{align}

    Indeed, consider $D,D'$ such that $D' = D + x$.
    \begin{itemize}
        \item 
    If $F = \emptyset$, $d \neq d_r$, or $e \not \in E_r$ we have $\rho_r^0(D) = \rho_r^0(D')$ for all such $D,D'$ so $\Delta_x(\rho_r^0) = 0$.
   \item If $F \neq \emptyset$, $d = d_r$ and $e \in E_r$ we have:
$\|\rho_r^0(D') - \rho_r^0(D) \|_1  \le \kappa$. 
        \end{itemize}

\end{proof}

\heading{Instantiation.}
    Side queries in both \Thm\ref{thm:bias_whole_report} and \Thm\ref{thm:bias_whole_report_last_touch}  follow the form from \Thm\ref{thm:bias_whole_report_sensitivity}, with $\rho_r^0(D) = \kappa \cdot \mathds{1}[\exists e \in E: D^e_{d_r} = \emptyset]$ in \Thm\ref{thm:bias_whole_report}  and $\kappa \cdot \mathds{1}[\exists i \in E_r: D^{i}_{d_r} = \emptyset \wedge \forall j \in E_r : j > i, D^j_{d_r} \cap F_A = \emptyset]$ in \Thm\ref{thm:bias_whole_report_last_touch}.
    Moreover, for these queries, the inequality in \Eq\ref{eq:sensitivity_count} is an equality if $|E| > 1$. For instance, consider $D = \{ (d_r, \hat e, F), \hat e \in E_r \setminus \{e\} \}, D' = \{ (d_r, \hat e, F), \hat e \in E_r \}$.
    This means that every requested device-epoch that has budget left and contains data should pay additional budget for the DP count.

\begin{theorem}[Sensitivity for certain histogram attribution functions]
    \label{thm:histogram_attribution_function}
    Consider an attribution function $A$ of the following form. 
    First, $A$ attributes a positive value $a_{\mathbf{F}}(f)$ to each relevant event $f \in F_1 \cap F_A \cup \dots \cup F_k \cap F_A$.
    Next, each event is mapped to a one-hot vector $H(f) \in \R^m$ (\ie $H(f) \in \{0,1\}^m$ and $\|H(f)\|_1 = 1$). 
    Finally, the attribution is the weighted sum:
    \begin{align}
    \label{eq:sum_over_events}
        A(F_1, \dots, F_k) = \sum_{i =1}^k \sum_{f \in F_i \cap F_A} a_{\mathbf{F}}(f) \cdot H(f)
    \end{align}

    We define:
    
    \begin{align}
        A^{\max} := \max_{\mathbf{F} \in \cP(\cI \cup \cC)^k }\sum_{i =1}^k \sum_{f \in F_i \cap F_A} a_{\mathbf{F}}(f)
    \end{align}

    Consider any attribution report $\rho_r$ with attribution function $A$ with output in $\R^m$. 

    \begin{itemize}
        \item If $m=1$ or $k = 1$, we have

        \begin{align}
        \Delta(\rho_r) \le \Delta^{\max}(\rho_r) \le A^{\max}
        \end{align}
        
        Moreover, if there exists $\mathbf{F}^{\max} = (\emptyset, \dots, \emptyset, \{f_0\}, \emptyset, \dots, \emptyset)$ containing a single relevant event $f_0 \in F_A$ such that $A^{\max}$ is attained, \ie $a_{\mathbf{F}^{\max}}(f_0) = A^{\max}$, then 

        \begin{align}
        \label{eq:m1_tight}
        \Delta(\rho_r) = \Delta^{\max}(\rho_r) = A^{\max}
        \end{align}

        \item If $m\ge2$ and $k\ge 2$, we have:
        \begin{align}
        \Delta(\rho_r) \le \Delta^{\max}(\rho_r) \le 2A^{\max}
        \end{align}

        Moreover, 
        if there exists $\mathbf{F}^{\max} = (\emptyset, \dots, \emptyset, \{f_0\}, \emptyset, \dots, \{f_1\}, \emptyset)$ and $\mathbf{F}'^{\max} = (\emptyset, \dots, \emptyset, \{f_0\}, \emptyset, \dots, \emptyset)$ such that $a_{\mathbf{F}^{\max}}(f_0) = A^{\max}$, $a_{\mathbf{F'}^{\max}}(f_1) = A^{\max}$ and $H(f_0) \neq H(f_1)$, then:

     \begin{align}
         \label{eq:m2_tight}
        \Delta(\rho_r) =  \Delta^{\max}(\rho_r) = 2A^{\max}
        \end{align}
    
    \end{itemize}

\end{theorem}
\begin{proof}
    Consider a report $\rho_r$ with such an attribution function $A$.
    First, we observe that $A(\emptyset) = 0 \in \R^m$, because of \Eq\ref{eq:sum_over_events}.
    
    We start by upper bounding $\Delta^{\max}(\rho_r)$.
    Take $\mathbf{F},\mathbf{F}' \in \cP(\cI \cup \cC)^k: \forall i \in [k], \mathbf{F}_i' = \mathbf{F}_i \text{ or } \mathbf{F}_i' = \emptyset$.

    \begin{itemize}
        \item If $m=1$, for any event $f$ we have $H(f) =1$. Since $a_{\mathbf{F}}(f) \ge 0$, we have:
        $ \sum_{i =1}^k \sum_{f \in F_i \cap F_A} a_{\mathbf{F}}(f) \cdot H(f) -\sum_{f \in F_i' \cap F_A} a_{\mathbf{F'}}(f) \cdot H(f) \le \sum_{i =1}^k \sum_{f \in F_i \cap F_A} a_{\mathbf{F}}(f) \cdot 1 \le A^{\max}$ and $\sum_{i =1}^k \sum_{f \in F_i \cap F_A} a_{\mathbf{F}}(f) \cdot H(f) -\sum_{f \in F_i' \cap F_A} a_{\mathbf{F'}}(f) \cdot H(f) \ge - \sum_{f \in F_i' \cap F_A} a_{\mathbf{F'}}(f) \cdot 1 \ge - A^{\max}$.
        Hence, $\|A(\mathbf{F}) - A(\mathbf{F}')\|_1 \le A^{\max}$, and thus $\Delta^{\max} \le A^{\max}$.

        \item If $k=1$, we have $\mathbf{F}' = F_1 \text{ or }  \emptyset $. In the first case, $\|A(\mathbf{F}) - A(\mathbf{F}')\|_1 = 0 \le A^{\max}$. In the second case, 
        \begin{align}
             \|A(\mathbf{F}) - A(\mathbf{F}')\|_1 &= \|A(\mathbf{F})\|_1 \\
             &\le\sum_{f \in F_1 \cap F_A} a_{\mathbf{F}}(f) \|H(f) \|_1\\ 
             &\le A^{\max}
        \end{align}
        Hence  $\Delta^{\max} \le A^{\max}$.
        
        \item If $m \ge 2$, we have:
            \begin{align}
       \|A(\mathbf{F}) - A(\mathbf{F}')\|_1 &= \|\sum_{i =1}^k \sum_{f \in F_i \cap F_A} a_{\mathbf{F}}(f) \cdot H(f)\\ 
       & \hspace{3mm} - \sum_{f \in F_i' \cap F_A} a_{\mathbf{F'}}(f) \cdot H(f)\|_1 \\
       &\le \sum_{i =1}^k \sum_{f \in F_i \cap F_A} a_{\mathbf{F}}(f) \|H(f) \|_1\\
       &+ \sum_{i =1}^k \sum_{f \in F_i' \cap F_A} a_{\mathbf{F'}}(f) \|H(f) \|_1\\
       &\le 2A^{\max}
    \end{align} 

           This is true for any such $\mathbf{F},\mathbf{F}'$, so $\Delta^{\max} \le 2A^{\max}$.
    
    \end{itemize}

    Next, we lower bound $\Delta^{\max}$. 
    \begin{itemize}
        \item  If  $m=1$ or $k=1$, and if there exists $\mathbf{F}^{\max} = (\emptyset, \dots, \emptyset, \{f_0\}, \newline \emptyset, \dots, \emptyset)$ such that $a_{\mathbf{F}^{\max}}(f_0) = A^{\max}$, we have
        \begin{align}
        \Delta^{\max}(\rho_r) &= \max_{\mathbf{F},\mathbf{F}' \in \cP(\cI \cup \cC)^k: \forall i \in [k], \mathbf{F}_i' = \mathbf{F}_i { or } \mathbf{F}_i' = \emptyset} \|A(\mathbf{F}) - A(\mathbf{F}')\|_1 \label{eq:m1_tight_exhibit} \\
        &\ge  \|A(\mathbf{F}^{\max}) - A(\emptyset)\|_1 \\
        &=  \| A^{\max} \cdot H(f_0) - 0 \|_1 \\
        &= A^{\max}
    \end{align}
    (in fact this is true even when $m\neq1$ and $k\neq1$). 
    \item If $m\ge 2$ and $k \ge 2$, and there exists $f_0, f_1$ such that removing $f_1$ shifts the attribution to $f_0$, and $H(f_0) \neq H(f_1)$, then:
    \begin{align}
        \Delta^{\max}(\rho_r) &\ge  \|A(\mathbf{F}^{\max}) - A(\mathbf{F}'^{\max})\|_1 \label{eq:m2_tight_exhibit} \\
        &=  \| A^{\max} \cdot H(f_0) - A^{\max} \cdot H(f_1) \|_1 \\
        &= 2A^{\max}
    \end{align}
    
    \end{itemize}

    We now focus on $\Delta(\rho_r)$. First, we have $\Delta(\rho_r) \le \Delta^{\max}(\rho_r)$, because if we note 
    $N := \{\mathbf{F},\mathbf{F}' \in \cP(\cI \cup \cC)^k :  \exists i \in [k]: \mathbf{F}'_i = \emptyset \wedge \forall j \neq i, \mathbf{F}'_i = \mathbf{F}_i\}$
    and 
    $N^{\max} := 
    \{
    \mathbf{F},\mathbf{F}' \in \cP(\cI \cup \cC)^k : 
    \forall i \in [k], \mathbf{F}'_i = \mathbf{F}_i \text{ or } \mathbf{F}'_i = \emptyset
    \}
    $
    we have $N \subset N^{\max}$.

    Second, the pairs of databases $\mathbf{F}^{\max}, \mathbf{F}'^{\max}$ exhibited in \Eq\ref{eq:m1_tight_exhibit} and \Eq\ref{eq:m2_tight_exhibit} happen to belong to both $N$
    and $N^{\max}$, so the upper bounds hold. 
\end{proof}

\heading{Instantiation.}
In particular, the upper bounds from \Thm\ref{thm:histogram_attribution_function} apply when the attribution function $A$ distributes a predetermined conversion value across impressions (\eg last-touch, first-touch, uniform, etc.), maps each impression to a bin (\eg $H(f)$ is a one-hot encoding of one of $m$ campaign identifiers), and then sums up the value in each bin. The resulting report $\rho_r(D) \in \R^m$ contains a histogram of the total attributed conversion value per bin.

The first tightness result (\Eq\ref{eq:m1_tight}) applies if there exists an impression that can be fully attributed. 
The second tightness result (\Eq\ref{eq:m2_tight}) applies if there exists two impressions $f_0, f_1$ with different one-hot encodings, such that removing $f_1$ shifts the maximum attribution value $A^{\max}$ to $f_0$ (\eg in last-touch attribution).

Note that we allow $A^{\max}$ to have any value, and we don't require every database to be fully attributed. This is a slight generalization of \cite{DGK+23}, which defines an {\em attribution rule} that requires $\sum_{i =1}^k \sum_{f \in F_i \cap F_A} a_{\mathbf{F}}(f) = 1$.